\documentclass[12pt]{iopart}
\usepackage{iopams}
\expandafter\let\csname equation*\endcsname\relax
\expandafter\let\csname endequation*\endcsname\relax
\usepackage{graphicx}
\usepackage{booktabs}
\usepackage{amssymb,bm,mathrsfs,bbm,amscd}
\usepackage[tbtags]{amsmath}

\begin{document}

\title{Relativistic tidal effects on clock-comparison experiments}

\author{Cheng-Gang Qin, Yu-Jie Tan, and Cheng-Gang Shao}

\address{MOE Key Laboratory of Fundamental Physical Quantities Measurements $\&$
  Hubei Key Laboratory of Gravitation and Quantum Physics, PGMF and School of Physics,
 Huazhong University of Science and Technology, Wuhan 430074, P.R. China}

\ead{{yjtan@hust.edu.cn,cgshao@hust.edu.cn}}\vspace{10pt}
\begin{indented}
\item[] October 23, 2018
\end{indented}

\begin{abstract}
We consider the relativistic tidal effects on frequency shift of clock-comparison experiments. The relativistic formulation for frequency shift and time transfer is derived in the gravitational field of a tidal, axisymmetric, and rotating Earth. With the help of Love numbers describing the tidal response of solid Earth, we formulize the mathematical connection between tidal effects from the ground-based clock-comparison experiments and the local gravity tides from the gravimeters, which in turn provides us an approach to eliminate tidal influences on clock comparison with the local gravity tides data. Moreover, we develop a method of the perturbed Kepler orbit to determine relativistic effects of clock comparison for space missions, which allows more precise calculations comparing to the conventional method of unperturbed Kepler orbit. With this perturbed method, it can give the perturbation of relativistic effects due to the orbital changes under the influences of tidal forces, Earth's oblateness etc.
In addition, as the applications of our results, we simulate tidal effects in frequency shift for the clock comparison on the ground and also give some estimates for TianQin mission and GPS.
\end{abstract}

\vspace{2pc} \noindent{\it Keywords}: frequency shift, clock comparison, gravitational redshift, relativistic tidal effects, perturbed Kepler problem

\section{Introduction}

Atomic clocks have been improved rapidly over the past years. The continued advances in clock stability and accuracy go hand in hand. At present, optical lattice clocks based on petahertz($10^{15}$Hz) transitions in isolated atoms are reaching stability and accuracy of order $10^{-18}$ in fractional frequency \cite{atomic1,atomic2,optical1,optical2,optical3,optical5,optical7}. The achieved accuracy has already asked for considering a new definition of the unit of time scales \cite{si1,si2,si3,si4}. The time keeping precision and frequency metrology precision at $10^{-18}$ could be utilized for testing relativity and standard model \cite{relativity1,relativity2,relativity3,relativity4,relativity5,relativity6,relativity8}, exploring new physical effects \cite{new2,new3,new4}, making the detection of gravitational waves \cite{wave1,wave2,wave3}, investigating possible variations of fundamental constants \cite{funda1,funda2,funda3}, enabling orders-of-magnitude improvements in geodesy and technology etc \cite{tech1,tech2}. For example, the first field measurement campaign with a transportable $^{87}$Sr optical clock demonstrates the exciting prospects for transportable optical clocks in geodesy and metrology \cite{transclock1,transclock2,transclock3}. Approaching $10^{-18}$ uncertainty optical atomic clocks in conjunction with optical links can serve as accurate measuring devices for the geopotential with several centimeters height resolution, which demonstrates a potential of achieving ``real-time" relativistic geodesy \cite{geodesy1,geodesy2,geodesy3,geodesy4}.

At the level of uncertainty of $10^{-18}$, a fully relativistic treatment of frequency comparison for the experiments in the vicinity of Earth must be performed up to the order $c^{-4}$ or $10^{-19}$.
Gravitational corrections of order $c^{-2}$ in the frequency shift has been determined a long time ago by Vessot $\emph{et al}$ \cite{vessot}.
Linet $\emph{et al}$ \cite{rel1} farther developed the relativistic frequency shift in the field of an axisymmetric rotating body.  Moreover, the interesting gravitomagnetic clock effect has been determined by Bini $\emph{et al}$ \cite{bini0,bini1}, who also treated some novel clock effects in Kerr spacetime \cite{bini2,bini3}. Recently, Derevianko $\emph{et al}$ \cite{dark1,dark2} proposed to search for dark matter in clock effects by the Global Positioning System (GPS) or network of atomic clock.
However, the relativistic tidal effects in frequency shift have not been determined with explicit calculations. It is the main motivation for this work to give a relativistic formulation for frequency shift and time transfer in the first post-Newtonian approximation for the gravitational field of a tidal, axisymmetric, and rotating Earth.

The conventional means of frequency transfer separated by thousands of kilometres is based on radiofrequency transfer via satellites, using GPS or a dedicated two-way satellite transfer \cite{fret1,fret2,fret3,fret4}. With the dissemination of microwave clock signals, their instability can reach $10^{-16}$ after one day measurement time. For near Earth missions, the Space-Time Explorer and QUantum Equivalence Space Test (STE-QUEST) space mission is expected to compare ground clocks down to the $1\times 10^{-18}$ uncertainty level with the microwave link \cite{relativity6},
where general relativistic effects are significant. For the gravitational effects of Earth, relativistic corrections due to the mass, quadrupole moment and intrinsic angular momentum have been studied \cite{rel1,ad5}, while the tidal effects due to Moon and Sun has not been solved yet. To account for tidal effects on satellite clocks, it requires to consider the contribution from two aspects. One is the contribution from the tidal potential itself, which directly causes frequency shift for an on-board clock. The other one is the contribution of similar orders of magnitude from changes in both position and velocity due to tidal perturbing forces ( the changes in both position and velocity leads to perturbations in the Earth's redshift and second-order Doppler effect.). Given a space mission in the vicinity of the Earth, the effects due to the tidal gravitational potentials induced by Moon and Sun need to be particularly treated with explicit calculations.

Recently, optical fibre links have shown the potential to transfer frequency with much better stability and accuracy \cite{fibre1,fibre2}. Optical fibre links over hundreds of kilometers are being developed worldwide and several groups around world have demonstrated short to medium distances ($10^{3}$ km) optical fibre links with the performance below $10^{-18}$ in frequency transfer \cite{link1,link2,link3}.
For optical fibre links, the fluctuations of the optical path due to the fibre temperature fluctuations and mechanical vibrations need to be compensated. An important technique is the "Doppler cancellation scheme" where the light makes a round trip in the same optical fibre in order to cancel or reduce some sources of error so that the measurement uncertainty is somewhat reduced. It also can be used to enhance corrections signal at the input end if the forward single trip correction is half of the total round-trip corrections.
For the increasing number of optical clocks in Europe, USA and Japan, a precise relativistic modeling of frequency comparison has to be studied where we should consider all factors involving frequency shift $10^{-18}$ or even $10^{-19}$. Except for influences of Earth, the influence of tidal potentials will be significant, which may reach several $10^{-17}$. It may be measured by the experiments of distant clocks comparison. The second-order tidal gravitational redshift due to Sun and Moon have semidiurnal and diurnal variations, which should be eliminated for many clock experiments in laboratories.   The third-order tidal gravitational redshift due to Moon also is a nonnegligible term since it can reach the level of $10^{-19}$. Since the Earth is not perfectly rigid, the tidal forces cause the deformation of Earth's shape and the redistribution of Earth mass. It in turn leads to changes in the Earth's gravitational potential, so the tidal response of solid Earth also needs to be taken into account.

In this work, we present a systematic, relativistic description of frequency comparisons for experiments in the vicinity of the Earth and give relativistic corrections up to the order $10^{-19}$. In Sec.\ref{II} we briefly introduce space-time reference frames and frequency shift formalism. In Sec.\ref{III}, we focus on the frequency comparisons in the terrestrial surface, in which the effects due to tidal potentials are important parts in the experiments. The Love numbers provide us a method to study tidal response of solid Earth \cite{snreio,love}. Tidal potential expansion in the celestial coordinate system allows us to conveniently analyze the characters of tidal effects. In Sec.\ref{IV} we use the time transfer function(TTF) formalism to determine frequency shift and time transfer in the gravitational field of a tidal, axisymmetric and rotating Earth. As an example for space missions of high orbiting spacecraft, we estimate the corresponding relativistic effects and Sagnac effect for TianQin. Then, we discuss the influences of tidal effects in Sec.\ref{V}, where the perturbed Kepler orbit is used to calculate the tidal effects on satellite clocks. In Sec.\ref{VI}, we give our conclusion. \ref{timetf} introduces briefly the time transfer function formalism. In \ref{apptidal}, we describe the tidal response of solid Earth through Love numbers. The perturbed Kepler problem is presented in \ref{appa}.

In this paper, $G$ is the Newtonian gravitational constant, and $c$ is the speed of the light in vacuum. The signature of the Lorentzian metric $g$ of space-time $V_{4}$ is given as $\{+---\}$. Greek indices run from 0 to 3 and Latin indices run from 1 to 3. We employ the vector notation $\textbf{a}$ in order to denote $(a^{1},a^{2},a^{3})=(a^{i})$. The Einstein convention on repeat indices is used here for the expressions like $a^{i}b^{i}$ and $A^{\mu}B_{\mu}$. The quantity $|\textbf{a}|$ stands for the ordinary Euclidean norm of $\textbf{a}$.
\section{Principles of relativistic frequency comparison}\label{II}
\subsection{Space-time reference frames}
The Einstein general relativity is generally covariant. In the Riemannian geometry, coordinate charts are merely labels. There is a wide freedom in the choice of coordinate system to describe the result of particular experiments. The space-time coordinates have no direct physical meaning and the physical observables are coordinate independent. Some available coordinate systems have important practical advantages in describing these observables of experiments. These coordinate systems are usually associated with a particular celestial body, laboratory or experiment facility. Using the harmonic gauge conditions and conservation laws, the relativistic proper reference frame associated with the particular body can be determined. In the DSX (Damour, Soffel $\&$ Xu) formalism, two useful coordinate reference systems (RSs) are barycentric and geocentric RSs \cite{dsx1,dsx2}.

The Solar System barycentric coordinate reference system (BCRS) is defined by IAU Resolution B1.3 (2000) \cite{gcrs1}.
It is a particular implementation of a barycentric reference system in the Solar System and has its origin at the Solar System barycenter. The BCRS is defined with the coordinates ${x^{\mu}_{B}}\equiv(ct_{B},\textbf{x}_{B}),$ with $t_{B}$ the barycentric coordinate time. The metric components of the BCRS may be written as follows \cite{gcrs1,ad1}:
\begin{eqnarray}\label{metricb}
  g^{B}_{00}=1-\frac{2w_{B}}{c^{2}}+\frac{2\beta w^{2}_{B}}{c^{4}}+O(c^{-6}),\nonumber\\
  g^{B}_{0i}=\frac{2(\gamma +1)\delta_{ik}w^{k}_{B}}{c^{3}}+O(c^{-5}),\\
  g^{B}_{ij}=-\delta_{ij}-\delta_{ij}\frac{2\gamma w_{B}}{c^2}-\frac{3\epsilon\delta_{ij}}{2}\frac{w^{2}_{B}}{c^4}+O(c^{-6}),\nonumber
\end{eqnarray}
where $\gamma$, $\beta$ and $\epsilon$ are the post-Newtonian parameters(in the general relativity, $\gamma=\beta=\epsilon=1$), $w_{B}$ and $\textbf{w}_{B}$ are the scalar and vector harmonic potentials, respectively. The scalar and vector potentials are mainly formed as a linear superposition of the gravitational potential contributed by the celestial bodies at the Solar System.

IAU Resolution B1.3 (2000) also defines the geocentric coordinate reference system (GCRS) \cite{gcrs1}. GCRS is a useful coordinate system in the vicinity of the Earth, which has its origin in the Earth center of mass.  Although BCRS could be used for all experiments, GCRS is physically adequate to the experiments which are implemented in the vicinity of the Earth. We can denote the coordinates of GCRS as ${x^{\mu}_{E}}\equiv(ct_{E},\textbf{x}_{E})$. The metric components of the GCRS may be written in the following form \cite{gcrs1,ad1}:
\begin{eqnarray}\label{metricg}
  g^{E}_{00}=1-\frac{2w_{E}}{c^{2}}+\frac{2\beta w^{2}_{E}}{c^{4}}+O(c^{-6}),\nonumber\\
  g^{E}_{0i}=\frac{2(\gamma +1)\delta_{ik}w^{k}_{E}}{c^{3}}+O(c^{-5}),\\
  g^{E}_{ij}=-1-\delta_{ij}\frac{2\gamma w_{E}}{c^2}-\frac{3\epsilon\delta_{ij}}{2}\frac{w^{2}_{E}}{c^4}+O(c^{-6}),\nonumber
\end{eqnarray}
where $w_{E}$ and $\textbf{w}_{E}$ are the scalar and vector harmonic potentials, respectively, and may be written as follows:
\begin{eqnarray}\label{3}
  w_{E}=U_{E}+u^{tidal}+O(c^{-4}),
\end{eqnarray}
and
\begin{eqnarray}\label{3evectorw}
  \textbf{w}_{E}=-\frac{GM_{E}}{2r_{E}^{3}}(\textbf{x}_{E}\times\textbf{s}_{E})+O(c^{-2}),
\end{eqnarray}
where $GM_{E}$ is the gravitational constant of Earth, $r_{E}=|\textbf{x}_{E}|$, the scalar potential $w_{E}$ is formed as a linear superposition of the gravitational potential $U_{E}$ of the isolated Earth and the tidal potential $u^{tidal}$ produced by all the Solar System bodies (excluding the Earth), the vector potential $\textbf{w}_{E}$ comes from the contribution of the Earth's rotation and $\textbf{s}_{E}$ is the Earth's angular momentum per unit of mass. $U_{E}$ is conveniently expressed in the form
\begin{equation}\label{4}
U_{E}(\textbf{x}_{E})= \frac{GM_{E}}{r_{E}}\Big{\{}1+\sum^{\infty}_{l=2}\sum^{+l}_{k=0}\Big{(}\frac{R_{0E}}{r_{E}}\Big{)}^{l}P_{lk}(\cos \theta)
 (C_{lk}\cos(k\phi)+S_{lk}\sin(k\phi))\Big{\}}+O(c^{-4}),
\end{equation}
where $R_{0E}$ is the Earth's equatorial radius, $P_{lk}$ are the Legendre polynomials. $C_{lk}$ and $S_{lk}$ relate to approximately constant potential coefficients in the terrestrial system corotating with Earth, which can be evaluated by numerically fitting to various kinds of experimental data, such as satellite motion tracking, geodetic measurements and gravimetry etc .
The tidal potential $u^{tidal}(\textbf{x}_{E})$ primarily comes from the contributions of the Moon and Sun. It can be given as usual:
\begin{eqnarray}\label{tidee}
  u^{tidal}=\sum_{b\neq E}(U_{b}(\textbf{r}_{bE}+\textbf{x}_{E})-U_{b}(\textbf{r}_{bE})-\textbf{x}_{E}\cdot \nabla U_{b}(\textbf{r}_{bE})),
\end{eqnarray}
where $U_{b}$ is the gravitational potential of the body $b$ (we mainly consider the Sun and Moon). $\textbf{r}_{bE}$ is the vector connecting the center of mass of body $b$ with that of Earth, and $\nabla U_{b}(\textbf{r}_{bE})$ is the gradient of the potential $U_{b}$.

From the assumption of the invariance of the Riemannian space-time interval $ds^{2}=g_{\mu\nu}dx^{\mu}dx^{\nu}$, the relationship between coordinate time and proper time is given by
\begin{eqnarray}\label{ttransfer}
  d\tau/dt=\sqrt{g_{00}+2g_{0i}p^{i}+g_{ij}p^{i}p^{j}},
\end{eqnarray}
where $\tau$ is the proper time, and $p^{i}=dx^{i}/cdt$ denotes the coordinate velocity. This equation allows to implement the time comparisons of the experiments in the Solar System. Based on Eq.(\ref{ttransfer}) and the metric tensor, the relationship between coordinate time and proper time of a clock can be expanded as follows:
\begin{equation}\label{ttransfer1}
  \frac{d\tau}{dt}=1-\frac{1}{2c^{2}}(\textbf{v}^{2}+2w)-\frac{1}{8c^{4}}[\textbf{v}^{4}+(4-8\beta)w^{2}+
  (4+8\gamma)w\textbf{v}^{2}-(16+16\gamma)\textbf{w}\cdot\textbf{v}]+O(c^{-6}),
\end{equation}
where $w$ and $\textbf{w}$ stand for the scalar and vector potentials in the specific reference system, respectively, and $\textbf{v}$ is the coordinate velocity. This formula is sufficient for the accuracy requirements of the all experiments in the next few years.

From one of recommendations of IAU Resolution (2000) \cite{gcrs1}, the barycentric coordinate time(TCB) and geocentric coordinate time(TCG) are defined as the time coordinates of BCRS and GCRS, respectively. The origin of GCRS is chosen at the Earth's center of mass so that the transformations of spatial coordinates and time coordinates don't include the effect of Earth's gravitational field. According to Eq.(\ref{ttransfer1}), the relationship between TCG and TCB is given by \cite{ad1}:
\begin{eqnarray}\label{tcg}
  \Big{\langle}\frac{dt_{E}}{dt_{B}}\Big{\rangle}=1-L_{C},
\end{eqnarray}
where $\langle\rangle$ refers to a sufficiently long average taken at the geocenter and $L_{C}=1.48082686741\times10^{-10}\pm2\times10^{-17}$ is a defined constant \cite{lc}, which is related to the matching problem between two systems. IAU Resolution B1.3 (2000) declares that the units of measurement of the coordinate times of BCGS and GCRS should be chosen so that they are consistent with the SI second. As the time scales of BCRS and GCRS, the relationship between TCB and TCG should be given by the time part of the full four-dimensional transformation between two reference systems. With the help of more precise observable and Eqs .(\ref{ttransfer1}) (\ref{tcg}), the constant $L_{C}$ can be theoretically determined in a more precise value after a long average.

\subsection{Frequency shift observables}
For calculating the observables of clocks comparison, the quantity of one-way frequency shift should be determined. We consider that a light signal of the frequency $\nu_{A}$ is emitted from the observer $O_{A}$ and the same signal is received by the observer $O_{B}$ at the frequency $\nu_{B}$. The one-way frequency shift between $O_{A}$ and $O_{B}$ can be expressed as follows\cite{ad5,ad4}:
\begin{eqnarray}\label{sp01}
   \frac{\nu_{B}}{\nu_{A}}=\frac{(u^{\mu}k_{\mu})_{B}}{(u^{\mu}k_{\mu})_{A}}=\frac{u^{0}_{B}k_{0}^{B}}{u^{0}_{A}k_{0}^{A}}\frac{1+p^{i}_{B}\hat{k}^{B}_{i}}{1+p^{i}_{A}\hat{k}^{A}_{i}},
\end{eqnarray}
where $u^{\mu}_{A}=(dx^{\mu}/ds)_{A}$ and $u^{\mu}_{B}=(dx^{\mu}/ds)_{B}$ are the four-velocity of $O_{A}$ and $O_{B}$, $p^{i}_{A}=(dx^{i}/cdt)_{A}$ and $p^{i}_{B}=(dx^{i}/cdt)_{B}$ are the coordinate velocities of $O_{A}$ and $O_{B}$, $k^{A}_{\mu}$ and $k^{B}_{\mu}$ are the null tangent vectors to the light signal at the point of emission $x_{A}$ and at the point of reception $x_{B}$, respectively.

TTF (see \ref{timetf}) formalism provides a direct way for determining the covariant components of the tangent vector to a photon trajectory\cite{ad5,ad2,ad4}
\begin{eqnarray}\label{ka}
 \hat{ k}^{A}_{i}=\frac{k^{A}_{i}}{k^{A}_{0}}=c\frac{\partial T_{r}}{\partial x^{i}_{A}}=-N^{i}_{AB}+\frac{\partial \Delta_{r}}{\partial x^{i}_{A}},
\end{eqnarray}
\begin{eqnarray}\label{kb}
 \hat{ k}^{B}_{i}=\frac{k^{B}_{i}}{k^{B}_{0}}
 =-\Big{(}N^{i}_{AB}+\frac{\partial \Delta_{r}}{\partial x^{i}_{B}}\Big{)}\times\Big{(}1-\frac{\partial \Delta_{r}}{c\partial t_{B}}\Big{)}^{-1},
\end{eqnarray}
\begin{eqnarray}\label{k0ab}
  \frac{k^{B}_{0}}{k^{A}_{0}}=1-\frac{\partial T_{r}}{\partial t_{B}}=1-\frac{\partial\Delta_{r}}{c\partial t_{B}},
\end{eqnarray}
where $\Delta _{r}$ is the ``delay function" or the gravitational delay, which is given by Eq.(\ref{ttfd2}) in the \ref{timetf}.

From the Eqs.(\ref{ka})-(\ref{k0ab}), the one-way frequency shift can be rewritten as follows:
\begin{eqnarray}\label{fretras2}
 \frac{\nu_{B}}{\nu_{A}}=\frac{(g_{00}+2g_{0i}p^{i}+g_{ij}p^{i}p^{j})^{1/2}_{A}}{(g_{00}+2g_{0i}p^{i}+g_{ij}p^{i}p^{j})^{1/2}_{B}}
  \frac{1-N^{i}_{AB}p^{i}_{B}-p^{i}_{B}\frac{\partial\Delta_{r}}{\partial x^{i}_{B}}-\frac{\partial\Delta_{r}}{c\partial t_{B}}}{1-N^{i}_{AB}p^{i}_{A}+p^{i}_{A}\frac{\partial\Delta_{r}}{\partial x^{i}_{A}}}.
\end{eqnarray}
However, there is another way to obtain the ratio of the covariant components of the tangent vector by the fact that it is equal to the ratio of the coordinate times \cite{ad6}
\begin{eqnarray}\label{sp001}
\frac{dt_{A}}{dt_{B}}=\frac{k^{B}_{0}(1+p^{i}_{B}\hat{k}^{B}_{i})}{k^{A}_{0}(1+p^{i}_{A}\hat{k}^{A}_{i})}.
\end{eqnarray}
We can calculate it directly by differentiating the coordinate time transfer $T_{AB}=t_{B}-t_{A}$ with respect to the received coordinate time $t_{B}$.

This modeling can be easily extended to a multi-way frequency shift. Let us consider a light signal emitted with the frequency $\nu_{A0}$ by observer $O_{A0}$, transmitted by observer $O_{A1}$, then transmitted again by observer $O_{A2}$... and finally received by observer $O_{An}$ with the frequency $\nu_{An}$. The frequency shift between $O_{A0}$ and $O_{An}$ is defined in the same way as for the one-way, which can be decomposed as follows:
\begin{eqnarray}\label{multif}
  \frac{\nu_{An}}{\nu_{A0}}=\frac{\nu_{An}}{\nu_{A(n-1)e}}\delta A_{(n-1)}...\delta A_{i}\frac{\nu_{Air}}{\nu_{A(i-1)e}}...\frac{\nu_{A1r}}{\nu_{A0}},
\end{eqnarray}
where $\nu_{Air}(i\in[1,n-1])$ is the proper frequency received by observer $O_{Ai}$, $\nu_{Aie}$ is the proper frequency emitted by the observer $O_{Ai}$, and $\delta A_{i}\equiv\nu_{Aie}/\nu_{Air}$ stands for the frequency shift contribution from the $i$th transponder, for example, the instrumental frequency shift due to fixed propagation paths of optical systems and the instability of instrument.
\section{The frequency shift in the terrestrial laboratory}\label{III}
Most of frequency-comparison experiments  are implemented in terrestrial laboratories. Optical fibers are widely applied for frequency and clock comparisons due to their superior performances. It requires us to study the relativistic corrections for the frequency transfer linked with optical fibres in terrestrial laboratories. We suppose that a light signal with proper frequency $\nu_{A}$ is emitted from observer $A$ at coordinate time $t_{A}$ and the same signal is received by observer $B$ with the proper frequency $\nu_{B}$ at coordinate time $t_{B}$. We another suppose that the signal travels in an optical fibre. The one-way frequency shift from $A$ to $B$ is characterized by $\nu_{B}/\nu_{A}$. From Eqs. (\ref{sp01}) and (\ref{sp001}), it can be written as
\begin{eqnarray}\label{15a}
  \frac{\nu_{B}}{\nu_{A}}=\frac{u^{0}_{B}}{u^{0}_{A}}\frac{dt_{A}}{dt_{B}}.
\end{eqnarray}
The derivative $dt_{A}/dt_{B}$ depends on the fibre itself, such as effective refractive index, and path of fibre $L$ (in this section and following sections we use GCRS and denote its coordinate as $x^{\mu}\equiv(ct,\textbf{x})$). This term includes influences of changing fibre length due to thermal expansion, changing refractive index or effect due to fibre motion. It should be determined for a optical fibre with lengths $\sim 1000$km to satisfy requirement of accuracy $10^{-19}$. Using the method in Ref.\cite{link1}, it can be expanded up to $c^{-3}$ order as follows (the error is far smaller than $10^{-19}$ for 1000km optical fibre):
\begin{eqnarray}\label{dtbdta}
  \frac{dt_{B}}{dt_{A}}=&&1+\frac{1}{c}\int^{L}_{0}\Big{(}\frac{\partial n}{\partial t}+n\alpha \frac{\partial T}{\partial t}\Big{)}dl
  -\frac{1}{c^{2}}\int^{L}_{0}\frac{\partial(\textbf{v}\cdot\textbf{s}_{l})}{\partial t}dl\nonumber\\
  &&+\frac{1}{c^{3}}\int^{L}_{0}\Big{\{}\frac{\partial }{\partial t}\Big{[(}w+\frac{\textbf{v}^{2}}{2}\Big{)}n\Big{]}+\Big{(}w+\frac{\textbf{v}^{2}}{2}\Big{)}n\alpha \frac{\partial T}{\partial t}\Big{\}}dl+O(c^{-4}),
\end{eqnarray}
where $l\in [0,L]$, $\textbf{v}=(\partial x^{i}/\partial t) \textbf{e}^{i}$, $\textbf{s}_{l}=(\partial x^{i}/\partial l) \textbf{e}^{i}$, $n$ is effective refractive index of the optical fibre, $\alpha$ is the linear thermal expansion coefficient of the optical fibre, and $T$ as a function of time and location is the temperature of the fibre. The $c^{-1}$ term corresponds to the the first-order Doppler effect. The $c^{-2}$ term is a time derivative of the Sagnac correction. The $c^{-3}$ term corresponds to the derivatives of the scalar potential and second-order Doppler correction, which can be neglected.

We consider the term of order $c^{-1}$ in detail. The changes of the effective refractive index and the temperature can be estimated on the experiments described in Refs.\cite{link1,ad7}. Their values can be estimated as $\partial n/\partial t\approx4\times10^{-11}$s$^{-1}$ and $\partial T/\partial t\approx4\times10^{-6}$K/s. The thermal thermal expansion coefficient is about $8\times10^{-7}$K$^{-1}$. The contribution of the $c^{-1}$-order term is about $1.5\times 10^{-19}$ for 1m long fibre.

We then consider the term $c^{-2}$ of Eq.(\ref{dtbdta}), which may be further expressed as \cite{link1}:
\begin{eqnarray}\label{sagnac2}
  \frac{1}{c^{2}}\int^{L}_{0}\frac{\partial\textbf{v}\cdot\textbf{s}_{l}}{\partial t}dl=\frac{1}{c^{2}}\int^{L}_{0}\textbf{a}\cdot\textbf{s}_{l}dl+\frac{1}{2c^{2}}(\textbf{v}^{2}_{B}-\textbf{v}^{2}_{A})
\end{eqnarray}
with $\textbf{a}=d\textbf{v}/d t$ the acceleration of points of the fibre.

The term $u^{0}_{B}/u^{0}_{A}$ doesn't depend on the fibre itself but on the states of the emitting and receiving observers. It includes the gravitational redshift and second-order Doppler effect. According to Eqs.(\ref{metricg}) and (\ref{ttransfer1}), this term can be written as
\begin{eqnarray}\label{groundfre1}
 \!\!\!\!\!\!\!\!\!\!\!\!\!\!\!\!\!\!\!\!\!\!\!\!\!\!\!\!\!\!\frac{u^{0}_{B}}{u^{0}_{A}}=1+\frac{1}{c^{2}}\Big{(}\frac{\textbf{v}^{2}_{B}}{2}-\frac{\textbf{v}^{2}_{A}}{2}+w_{B}-w_{A}\Big{)}
  +\frac{1}{c^{4}}\Big{\{}\frac{1}{2}(w_{B}-w_{A})(w_{B}+\textbf{v}_{B}^{2}-w_{A}-\textbf{v}_{A}^{2})\\
  \!\!\!\!\!\!\!\!\!\!\!\!\!\!\!\!-\frac{1}{8}(\textbf{v}_{A}^{4}+2\textbf{v}_{B}^{2}\textbf{v}_{A}^{2}-3\textbf{v}_{B}^{4})+2(w_{B}\textbf{v}_{B}^{2}-w_{A}\textbf{v}_{A}^{2})
  -4(\textbf{w}_{B}\cdot\textbf{v}_{B}-\textbf{w}_{A}\cdot\textbf{v}_{A})\Big{\}}+O(c^{-6}),\nonumber
\end{eqnarray}
where we take the values of post-Newtonian parameters as $\gamma=\beta=\epsilon=1$ (for these parameters, the limits from space missions and astrometric observables are much better than that from ground experiments, so it is sufficient to set $\gamma=\beta=\epsilon=1$ in laboratories ). The terms of order $c^{-2}$ contain the second-order Doppler effect and the Earth gravitational redshift effect, which must be taken into account in the following calculations.  For the terms of order $c^{-4}$, the second term does not exceed $10^{-24}$, which can be neglected. The contribution of the third term is smaller than $10^{-20}$ due to the coupling of the scalar potential and velocity. The fourth term is smaller than $10^{-20}$ due to the contribution of vector potential. Only reserving the first term of the order $c^{-4}$, $u^{0}_{B}/u^{0}_{A}$ may be rewritten as
\begin{equation}\label{groundfre2}
  \frac{u^{0}_{B}}{u^{0}_{A}}=1+\frac{1}{c^{2}}\Big{(}\frac{\textbf{v}^{2}_{B}}{2}-\frac{\textbf{v}^{2}_{A}}{2}+U^{B}_{E}-U^{A}_{E}+u^{tidal}_{B}-u^{tidal}_{A}\Big{)}
  +\frac{1}{2c^{4}}(U^{B}_{E}-U^{A}_{E})^{2}.
\end{equation}
Considering the accuracy requirement of modern experiments, the astronomical tidal potentials are nonnegligible as well as the tidal response of solid Earth. From a rough estimation, the contributions of the tidal potentials due to Moon and Sun can reach several $10^{-17}$. The tidal effects are significant for many frequency comparisons. With this motivation, we will study them with the LOVE numbers \cite{snreio,tid} in \ref{apptidal}.

The term $u^{0}_{A}$ and $u^{0}_{B}$ depend on the states of the emitting and receiving observers on the terrestrial surface, respectively. The influences of the tidal response of solid Earth and tidal gavitational potentials induced by the Sun and Moon have been represented in \ref{apptidal}. With the help of Eq.(\ref{5t}), Eq.({\ref{groundfre2}}) may be rewritten as
\begin{eqnarray}\label{16}
\frac{u^{0}_{B}}{u^{0}_{A}}=1+\frac{1}{c^{2}}\Big{[}\frac{\textbf{v}^{2}_{B}-\textbf{v}^{2}_{A}}{2}+(U_{E}(\textbf{x}_{B})-U_{E}(\textbf{x}_{A}))+(1-h_{2}+k_{2})\times \\
  (u^{tidal}_{2}(\textbf{x}_{B})-u^{tidal}_{2}(\textbf{x}_{A}))
 +(1-h_{3}+k_{3}) (u^{tidal}_{3m}(\textbf{x}_{B})-u^{tidal}_{3m}(\textbf{x}_{A}))\Big{]}+O(c^{-4}),\nonumber
\end{eqnarray}
where $\textbf{x}_{A}$ and $\textbf{x}_{B}$ are the positions of $A$ and $B$, respectively. For the term of order $c^{-2}$, the first term is a constant that depends on the two positions on the Earth's surface (due to the change of height, the change of that does not exceed $10^{-19}$ for the two positions at a distance of 1000km). The second term is also a constant that comes from the height difference of the two positions, in other way, the gravitational redshift due to Earth gravitational potential (the fluctuating part of Earth potential due to Earth deformation and the redistribution of mass has been put into the third and fourth terms). The third term comes from the difference of second-order tidal potentials between positions $A$ and $B$. After including contributions of tidal response of the solid Earth, the coefficient $1-h_{2}+k_{2}$ may be treated as a constant $0.7$ for clock comparisons at the level of accuracy $10^{-18}$. This term is crucial for clock or frequency comparisons in the laboratory, whose amplitude approximates $1\times10^{-17}$ for clock or frequency comparisons of distances $\sim 1000$km. The fourth term corresponds to the difference of the third-order Moon's tidal potentials between positions $A$ and $B$, since the third-order Sun's tidal potential is much less than the second-order Sun's tidal potential. The coefficient $1-h_{3}+k_{3}$ can be treated as a constant 0.8 for the level of accuracy $10^{-19}$. This term is crucial for clock or frequency comparisons in the laboratory, whose amplitude reaches $10^{-19}$ for clock or frequency comparisons of distances $\sim 1000$km.

On the Earth surface, it is convenient to rewrite Eq.(\ref{16}) as:
\begin{equation}\label{16a}
\frac{u^{0}_{B}}{u^{0}_{A}}=1+\frac{1}{c^{2}}\Big{[}\int gdh
 -\frac{R_{E}(1-h_{2}+k_{2}) }{2+2h_{2}-3k_{2}}(\delta g_{B}-\delta g_{A})\Big{]}+O(c^{-4})
\end{equation}
$g$ is the gravity acceleration, and $dh$ is the length increment along the positive upward plumb line. This formula is sufficient for all modern experiments in labs (accuracy $10^{-18}$), which provides us a formula to study the frequency shift arising from tidal potentials by the tidal data. To establish a more precision model(accuracy $10^{-19}$), we should use Eq.(\ref{16}).
\begin{figure}
  \centering
  \includegraphics[width=0.8\textwidth]{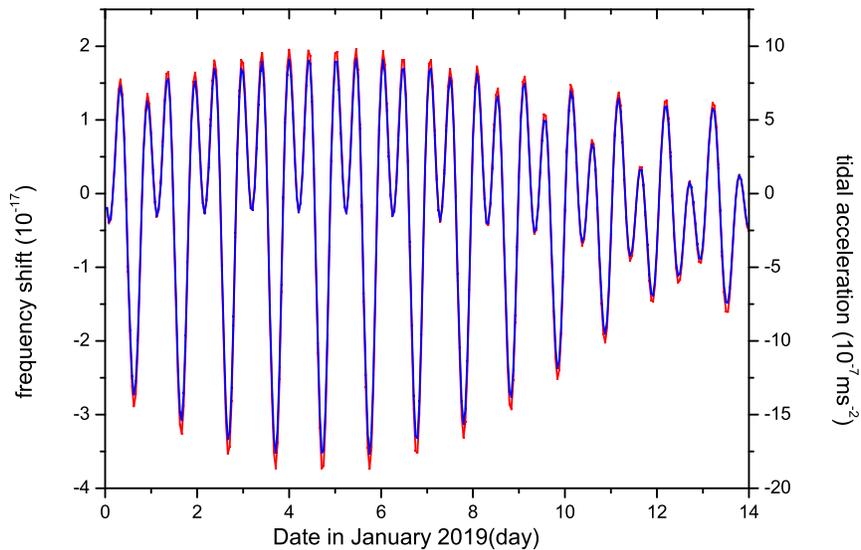}
  \caption{Frequency shift and local tidal acceleration on the position $A$(east longitude $114^{\circ}$, latitude $30^{\circ}$) for 14 days, from 1 January 2019. The left axis represents frequency shift(red line) due to the total tidal potentials, and right axis represents local tidal acceleration (blue line). Two curves are almost coincident with each other, but have different magnitudes.}\label{figure1}
\end{figure}
In order to demonstrate the contribution of the fluctuating term, we simulate the frequency comparison between two locations, $A$ and $B$. East longitude and latitude of $A$ are $114^{\circ}$ and $30^{\circ}$ (WuHan), and that of $B$ are $116^{\circ}$ and $40^{\circ}$(BeiJing), respectively. The distance of $A$ and $B$ is about 1100km. We shall use nominal planetary ephemeris, astronomy parameters$^{\footnotemark[1]}$ and general gravitational constants of Earth, Sun and Moon. The signals of the frequency shift and tides are given by simulation. Here we mainly focus on the contribution of the fluctuating term. The simulation time begins with January 1, 2019.
\footnotetext[1]{We give some astronomy parameters used in our work. Such as, the average distance $384400\times10^{3}$m between Earth and Moon, and average distance $149600\times10^{6}$m between Earth and Sun.}

\begin{figure}
  \centering
  \includegraphics[width=0.8\textwidth]{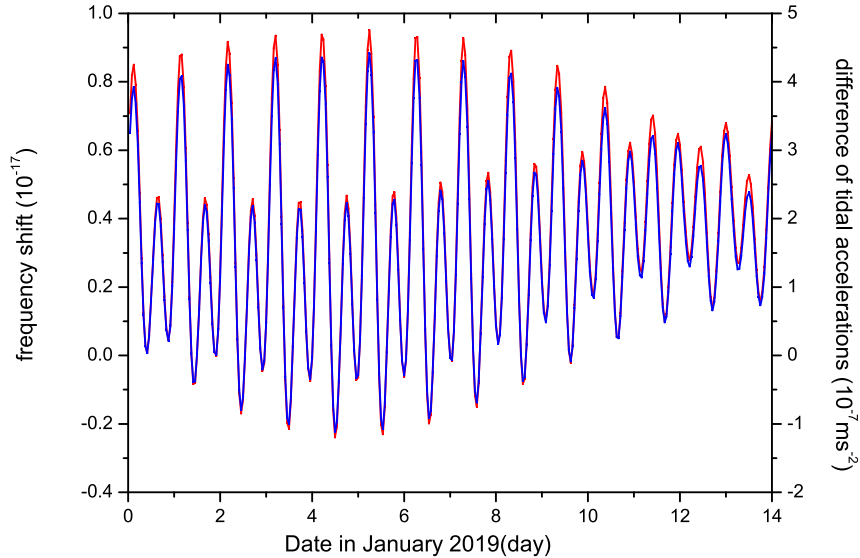}
  \caption{The calculated frequency shift and difference of tidal accelerations between $A$ and $B$ for 14 days. The distance between $A$ and $B$  is about 1000km. The red line and blue line represent frequency shift and difference of local tidal accelerations between $A$(east longitude $114^{\circ}$, latitude $30^{\circ}$) and $B$(east longitude $116^{\circ}$, latitude $40^{\circ}$), respectively. Their curves are consistent with each other, but have different magnitudes (left is $10^{-17}$ and right is $10^{-7}$ms$^{-2}$).}\label{figure2}
\end{figure}

Figure \ref{figure1} shows the simulations of tidal acceleration (tides data) of position $A$ and frequency shift contributed by the total tidal potentials(the astronomical tidal and solid Earth tidal potentials). The frequency shift (red line) is almost coincident with the local tidal acceleration (blue line), but just has different magnitudes. The total time is 14 days from 1 January 2019 to 14 January 2019. The fluctuating frequency shift can reach $4\times10^{-17}$. There has an obvious periodicity in Figure \ref{figure1}. This figure illustrates that tidal effects in frequency shift arising from astronomical tidal potentials, deformation of Earth and redistribution of Earth mass are proportional to local tidal acceleration (for example, gravity tides from superconducting gravimeters). So we could apply tides data to the primary frequency standard, geopotential and geodesy.

\begin{figure}
  \centering
 \includegraphics[width=0.8\textwidth]{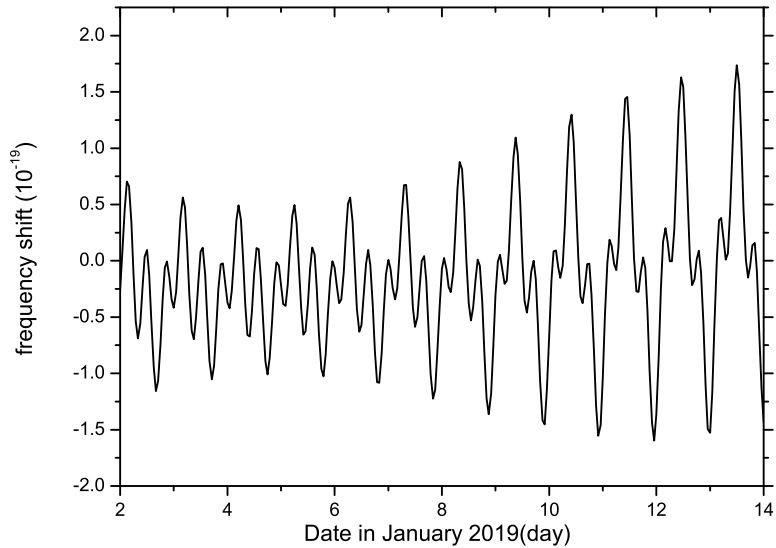}
  \caption{The calculated frequency shift for 14 days. Frequency comparison of the separated clocks is simulated between $A$ and $B$ at a distance of about 1000km. The line represents frequency shift due to the third-order tidal potentials. }\label{figure3}
\end{figure}

Figure \ref{figure2} shows the calculated frequency shift and difference of local tidal accelerations between $A$ and $B$ for 14 days. The red line is the frequency shift of comparing the separated clocks between $A$ and $B$ due to the total tidal potentials from 1 January 2019 to 14 January 2019. The blue line represents the difference of local tides accelerations. This figure illustrates the signal of frequency comparison is coincident with the signal of difference of tidal acceleration between $A$ and $B$, but they have different magnitudes. The frequency shift has an order of magnitude $10^{-18}$, and can reach $1\times10^{-17}$ for a distance of 1000km. The difference of tidal accelerations is several $10^{-7}$ms$^{-2}$. It is clear that the tidal gravitational redshift is significant and measurable in this case. The experiments should consider tidal effects for the frequency comparisons at the distances of several $100$km, and clock-comparison procedures need to rigorously consider influences of the astronomical tidal potentials and Earth response to the tidal forces. For the frequency comparisons by optical fiber link in laboratories, we can apply tidal acceleration data to remove the contributions due to the total tidal potentials. As an application, we may apply the local tides data of laboratories in the data of computing TAI in the future.

Figure \ref{figure3} shows the frequency shift due to the third-order tidal potentials between $A$ and $B$. The third-order tidal potentials consist of the third-order Moon tidal potential and third-order tidal response of solid Earth. This tidal contribution approximately amounts to $2\times10^{-19}$ for the frequency comparison. The third-order tidal frequency shift can reach $10^{-19}$ for clock comparisons at distances $\sim10^{3}$km. Different from Figure \ref{figure1}, it has the period of 1/3 day which may be measured in the future.
The periods in Figures \ref{figure1} and  \ref{figure3} are coincident with Eqs.(\ref{9}) and (\ref{tid3m2}), respectively. The tidal potential expressions in the celestial coordinate system may help us to better distinguish tidal effects and other periodic effects. A better understanding and extraction of tidal effects may help to study possible new physics.
\section{The frequency shift in space missions}
Different to experiments in the terrestrial laboratory having the fixed optical-fiber paths and positions in Earth, the experiments of space missions complicate calculation procedure. In space missions, the position and velocity of satellite (clock) are time-evolution, and the paths of light signal are dependent to satellite positions, thus the relativistic effects will depend on the orbital parameter. The formulae for terrestrial laboratory are not suitable for experiments of space missions, which implies we should reconsider the frequency sift. In order to make discussion more clear, we give the analysis in following subsection. The signal-path-dependent part ($q_{B}/q_{A}$) is mainly discussed by the case of high Earth-orbiting satellites. And the discussion of clock-dependent (${u^{0}_{B}}/{u^{0}_{A}}$) part is given by the GPS-like case.
\subsection{The frequency shift between high Earth-orbiting satellites}\label{IV}
In this section, we consider the frequency comparisons for the case of space to space. The frequency comparisons on this case have many applications in space missions. Many space missions use high orbit satellites to implement measurements, such as TianQin\cite{wave3} and the Beyond Einstein Advanced Coherent Optical Network(BEACON) mission\cite{be1,be2}. With the development of space technology, it has reached the accuracy of order $10^{-16}$ in fractional frequency, which also motivates us to study the frequency comparisons of this case. As an example, we perform numerical estimates of frequency shift for the TianQin, which is a space-borne gravitational wave detection mission relying on a constellation of three identical satellites. Those satellites will be placed on nearly identical geocentric orbits with semi-major axis of $\sim1\times10^{8}$m forming a nearly equilateral triangle. Each arm's length of the laser interferometer is about $1.73\times10^{8}$m. For the velocity of satellites, we take the value 2.0km/s. The parameters of Earth are $GM_{E}=3.986\times10^{14}$m$^{3}$s$^{-2}$, $R_{0E}=6.378\times10^{6}$m and $J_{2}=1.083\times10^{-3}$.

We assume two satellites $S_{A}$ and $S_{B}$ are moving on Earth's orbits. From Eq.(\ref{fretras2}), the one-way frequency shift from $S_{A}$ to $S_{B}$ is written as
\begin{eqnarray}\label{spacefre}
  \frac{\nu_{B}}{\nu_{A}}=\frac{u^{0}_{B}}{u^{0}_{A}}\frac{q_{B}}{q_{A}},
\end{eqnarray}
where
\begin{eqnarray}\label{spacefreu}
 \frac{u^{0}_{B}}{u^{0}_{A}}=1+\frac{1}{c^{2}}\Big{(}\frac{\textbf{v}^{2}_{B}}{2}-\frac{\textbf{v}^{2}_{A}}{2}+w_{B}-w_{A}\Big{)}
  +\frac{1}{c^{4}}\Big{\{}(1+\gamma)(w_{B}\textbf{v}_{B}^{2}-w_{A}\textbf{v}_{A}^{2})\nonumber\\
  +(1-\beta)(w^{2}_{B}-w^{2}_{A})+\frac{1}{2}(w_{B}-w_{A})^{2}+\frac{1}{2}(w_{B}-w_{A})(\textbf{v}_{B}^{2}-\textbf{v}_{A}^{2})+\frac{\textbf{v}_{B}^{4}}{2}\nonumber\\
  -\frac{1}{8}(\textbf{v}_{B}^{2}+\textbf{v}_{A}^{2})^{2}-2(1+\gamma)(\textbf{w}_{B}\cdot\textbf{v}_{B}-\textbf{w}_{A}\cdot\textbf{v}_{A})\Big{\}}+O(c^{-6})
\end{eqnarray}
and
\begin{eqnarray}\label{spacefreq}
  \frac{q_{B}}{q_{A}}=1-\textbf{N}_{AB}\cdot(\textbf{p}_{B}-\textbf{p}_{A})\sum^{3}_{n=0}(\textbf{N}_{AB}\cdot\textbf{p}_{A})^{n}
  -(1+\textbf{N}_{AB}\cdot\textbf{p}_{A})\nonumber\\
  \times\Big{(}p^{i}_{B}\frac{\partial\Delta_{r}}{\partial x^{i}_{B}}+p^{i}_{A}\frac{\partial\Delta_{r}}{\partial x^{i}_{A}}+\frac{\partial\Delta_{r}}{c\partial t_{B}}\Big{)}
+\textbf{N}_{AB}\cdot(\textbf{p}_{B}-\textbf{p}_{A})p^{i}_{A}\frac{\partial\Delta_{r}}{\partial x^{i}_{A}}+O(c^{-5}).
\end{eqnarray}
In order to obtain the frequency shift due to gravitational delay induced by tidal potential, we focus on the calculation of Eq.(\ref{spacefreq}) in this section. The more detailed calculation of Eq.(\ref{spacefreu}) is given by Sec.\ref{V}.

For Eq.(\ref{spacefreu}), a simple Kepler orbit is not sufficient to estimate all relativistic effects.  Since the perturding forces cause changes in both position and velocity of satellite, it would change the contribution of the monopole potential of earth to the satellite clock frequency.  The contributions of perturbing forces should be added in the formalism, such as, Moon's and Sun's tidal forces. Therefore, term $u^{0}_{B}/u^{0}_{A}$ may be rewritten as
\begin{equation}\label{simpleruab}
   \frac{u^{0}_{B}}{u^{0}_{A}}=1+\frac{1}{c^{2}}\Big{(}U^{B}_{E}-U^{A}_{E}+\frac{v^{2}_{B}}{2}-\frac{v^{2}_{A}}{2}
   +u^{B}_{tidal}-u^{A}_{tidal}\Big{)}+\delta_{per,B}-\delta_{per,A}+O(c^{-4}).
\end{equation}
For rough estimates, the monopole term in the Newtonian potential of the Earth produces a contribution $GM_{E}/c^{2}r\simeq 4.4\times10^{-11}$ for a TianQin spacecraft. The quadrupole term produces a contribution $GM_{E}J_{2}R_{0E}^{2}/c^{2}r^{3}\simeq 1.95\times10^{-16}$. The contributions of higher mass multipole terms are negligible. The contribution of the tidal potential term may reach the order of magnitude $10^{-14}$, since tidal potentials grow quadratically with distance from Earth's center of mass. The last two terms of Eq. (\ref{simpleruab}) are perturbing contributions due to the influence of perturbing forces in orbits. Each term can be split into two parts, the perturbation in gravitational redshift $(\delta f/f)_{Eper}$ and perturbation in second-order Doppler effect$(\delta v^{2}/c^{2})_{per}$, which can be expressed as Eqs. (\ref{eredper}) and (\ref{v2per}) (procedures are given in the next section and \ref{appa} ). As described in \ref{appa}, the projections of perturbing forces on the orbital coordinate system are $\mathcal{A},\mathcal{B}$ and $\mathcal{C}$. However, the components of perturbing forces are more complicated if we consider the perturbing forces for Sun's tidal force, Moon's tidal force, the oblateness of Earth etc. Component $\mathcal{A}$ can be written as the form
\begin{eqnarray}\label{aperall}
  \mathcal{A}=\mathcal{A}_{m}+\mathcal{A}_{s}+\mathcal{A}_{J_{n}}+...
\end{eqnarray}
where $\mathcal{A}_{m}$ represents contribution of the Moon, $\mathcal{A}_{s}$ represents contribution of the Sun, $\mathcal{A}_{J_{n}}$ represents contribution of the oblateness of Earth, and ellipsis represents the contributions of other forces. The other components $\mathcal{B}$ and $\mathcal{C}$ can be written with the similar forms. For the contributions of Moon, the perturbation may reach several parts in $10^{15}$.

For the right side of $q_{B}/q_{A}$, the second term is the special relativity Doppler effect, which is just related to the velocities of two satellites. The third and fourth terms include the influences of the gravitational field. By solving the delay function up to the order $c^{-3}$, the frequency shift can be determined to the order $c^{-4}$. With the conception of TianQin, it is easy to deduce the bounds: $\textbf{N}_{AB}\cdot\textbf{p}_{A/B}\leq6.7\times10^{-6}$ for one satellite and $\textbf{N}_{AB}\cdot(\textbf{p}_{B}-\textbf{p}_{A})\leq1.34\times10^{-5}$ for the first-order Doppler effect. The case $n\geq3$ for the second term can be neglected.

In order to calculate the third and fourth terms of $q_{B}/q_{A}$, the ``delay function" (or gravitational delay) $\Delta_{r}$ remains to be solved. With the help of times transfer function formalism in the \ref{timetf}, ``delay function" $\Delta_{r}$ can be solved. It may be written as follows:
\begin{eqnarray}\label{rewdelay}
  \Delta_{r}=\Delta_{M_{E}}+\Delta_{tidal}+\sum^{\infty}_{n=2}\Delta_{J_{n}}+\Delta_{s}+O(c^{-4})
\end{eqnarray}
where these four terms are deduced from $\Delta^{(1)}_{r}$ of Eq.(\ref{generalfun}). $\Delta_{M_{E}}$ represents the contribution of the mass monopole, $\Delta_{J_{n}}$ represents the contributions from the mass multipole moments, and $\Delta_{s}$ represents the contribution of the vector potential. Since these three terms have been determined by some works\cite{shapiro,ad12,ad14}, we will direct use the previous results to the calculations for frequency shift. Here, we will focus on the second term $\Delta_{tidal}$.

The second term $\Delta_{tidal}$ of Eq.(\ref{rewdelay}) represents the contributions from the tidal potentials, which is given by the integral along the signal path,

\begin{equation}\label{tidefull01}
  \Delta_{tidal}=\frac{(1+\gamma)R_{AB}}{c^{2}}\int^{1}_{0}\sum_{b\neq E} \Big{[}\frac{GM_{b}}{|\textbf{r}_{bE}+\textbf{x}_{B}-\lambda \textbf{R}_{AB}|}-\frac{GM_{b}}{|\textbf{r}_{bE}|}+(\textbf{x}_{B}-\lambda \textbf{R}_{AB})\cdot\textbf{r}_{bE}\frac{GM_{b}}{r^{3}_{bE}} \Big{]} d\lambda.
\end{equation}
Taking this equation into account, we obtain
\begin{equation}\label{tidefull1}
  \Delta_{tidal}=\frac{(1+\gamma)R_{AB}}{c^{2}}\sum_{b\neq E}\Big{[}\frac{GM_{b}}{R_{AB}}\ln\frac{r_{bA}+r_{bB}+R_{AB}}{r_{bA}+r_{bB}-R_{AB}}-\frac{GM_{b}}{r_{bE}}
  +\frac{GM_{b}}{r^{3}_{bE}}\frac{(\textbf{x}_{A}+\textbf{x}_{B})\cdot\textbf{r}_{bE}}{2}\Big{]},
\end{equation}
where $r_{bA}=|\textbf{x}_{A}+\textbf{r}_{bE}|$, $r_{bB}=|\textbf{x}_{B}+\textbf{r}_{bE}|$ . This equation allows us to calculate tidal effects due to all bodies (except for Earth) in the Solar System. In general, it is sufficient to consider the second-order and third-order tidal potentials due to Moon and Sun for modern space missions. With the help of expansion of tidal potential Eq.(\ref{6}), it is reasonable to express the tidal term as $\Delta_{tidal}=\Delta_{tidal2}+\Delta_{tidal3}$. After some calculations, we obtain
\begin{equation}\label{ttftid2}
  \Delta_{tidal2}=\sum_{b\neq E}\frac{(1+\gamma)GM_{b}R_{AB}}{4c^{2}r_{bE}^{3}}\Big{[}6(\textbf{n}_{bE}\cdot\textbf{x}_{B})(\textbf{n}_{bE}\cdot\textbf{x}_{A})
  +2(\textbf{n}_{bE}\cdot\textbf{R}_{AB})^{2}-2\textbf{x}_{B}\cdot\textbf{x}_{A}-\frac{2}{3}R^{2}_{AB}\Big{]},
\end{equation}
and
\begin{eqnarray}\label{ttftid3}
  \!\!\!\!\!\!\!\!\!\!\!\!\!\!\!\!\!\!\!\!\!\Delta_{tidal3}=\sum_{b\neq E}\frac{(1+\gamma)GM_{b}R_{AB}}{4c^{2}r_{bE}^{4}}\Big{[}15(\textbf{n}_{bE}\cdot\textbf{x}_{B})(\textbf{n}_{bE}\cdot\textbf{x}_{A})^{2}
  +\frac{5}{2}(\textbf{n}_{bE}\cdot\textbf{R}_{AB})^{3}\nonumber\\
  \!\!\!\!\!\!\!\!\!\!\!\!\!\!\!\!\!\!\!\!\!+10(\textbf{n}_{bE}\cdot\textbf{R}_{AB})^{2}(\textbf{n}_{bE}\cdot\textbf{x}_{A})-5(\textbf{n}_{bE}\cdot\textbf{x}_{A})^{3}
  -\frac{3}{2}R^{2}_{AB}(\textbf{n}_{bE}\cdot\textbf{x}_{B})-\frac{1}{2}R^{2}_{AB}(\textbf{n}_{bE}\cdot\textbf{x}_{A})\nonumber\\
  \!\!\!\!\!\!\!\!\!\!\!\!\!\!\!\!\!\!\!\!\!-4(\textbf{R}_{AB}\cdot\textbf{x}_{A})(\textbf{n}_{bE}\cdot\textbf{R}_{AB})
 -6(\textbf{x}_{B}\cdot\textbf{x}_{A})(\textbf{n}_{bE}\cdot\textbf{x}_{A})-3(\textbf{n}_{bE}\cdot\textbf{R}_{AB})\textbf{x}^{2}_{A}\Big{]}
\end{eqnarray}
with $r_{bE}=|\textbf{r}_{bE}|$ and $\textbf{n}_{bE}=\textbf{r}_{bE}/r_{bE}$. Eq.(\ref{ttftid2}) has to take Moon and Sun into account. Since the third-order Sun's tidal potential is much less than second-order tidal potential, it is sufficient to only calculate Moon's contribution for Eq.(\ref{ttftid3}). For low Earth-orbiting satellites, it is sufficient to use Eqs.(\ref{tidefull1})-(\ref{ttftid3}). However, what we need to point out is that Eq.(\ref{tidefull1}) is more suitable than Eqs.(\ref{ttftid2}) and (\ref{ttftid3}) for high Earth-orbiting satellites. $\Delta_{tidal}$ can reach the level of $10^{-6}$m for TianQin space mission.

Differentiating Eq.(\ref{tidefull1}) with respect to $\textbf{x}_{A}$ and $\textbf{x}_{B}$, we can obtain the frequency shift due to tidal contributions on the propagation of light, which is
\begin{eqnarray}\label{dtidefull1ab}
    \!\!\!\!\!\!\!\!\!\!\!\!\!\!\!\!\!\!\!\!\!\!\!\!\!\!\!\Big{(}\frac{\delta\nu}{\nu}\Big{)}_{\Delta_{tidal}}=\sum_{b\neq E}
  \frac{(1+\gamma)GM_{b}}{c^{3}}
  \Big{\{}\frac{R_{AB}(\textbf{n}_{bA}\cdot\textbf{v}_{A}+\textbf{n}_{bB}\cdot\textbf{v}_{B})-(r_{bA}+r_{bB})(\textbf{N}_{AB}\cdot\textbf{v}_{AB})}{r_{bA}r_{bB}+\textbf{r}_{bA}\cdot\textbf{r}_{bB}}\nonumber\\
    \!\!\!\!\!\!\!\!\!\!\!\!\!\!\!\!\!\!\!\!\!\!\!\!\!\!\!+\frac{\textbf{N}_{AB}\cdot\textbf{v}_{AB}}{r_{bE}}-
  \frac{[(\textbf{x}_{A}+\textbf{x}_{B})\cdot\textbf{x}_{bE}](\textbf{N}_{AB}\cdot\textbf{v}_{AB})+R_{AB}[\textbf{x}_{bE}\cdot(\textbf{v}_{A}+\textbf{v}_{B})]}{2r_{bE}^{3}} \Big{\}}
\end{eqnarray}
with $\textbf{v}_{AB}=\textbf{v}_{B}-\textbf{v}_{A}$, and $\textbf{N}_{AB}=\textbf{R}_{AB}/R_{AB}$. Taking Moon as an example, the Moon's contributions may reach the level of $10^{-19}$ for the case $\gamma=1$. To make the discussion more convenient, it is useful to calculate contributions of the second-order tidal potentials. From the Eq.(\ref{ttftid2}), it is easy to obtain derivatives of the second-order tidal term.
Inserting these derivatives into Eq.(\ref{spacefreq}), the frequency shift due to gravitational delay of the second-order tidal potentials can be determined. Since the tidal corrections is negligible for tidal term, it is sufficient for estimate with Keplerian orbit. Neglecting terms of $e^{2}$, the frequency shift due to the second-order tidal contributions on the propagation of light is given by
\begin{eqnarray}\label{dttftidab}
 \!\!\!\!\!\!\!\!\!\!\!\!\!\!\!\!\!\!\!\!\!\!\!\!\!\!\!\Big{(}\frac{\delta\nu}{\nu}\Big{)}_{\Delta_{tidal2}}
  =\sum_{b\neq E}\frac{(1+\gamma)GM_{b}}{4c^{3}r_{bE}^{3}}\Big{\{}\sqrt{3}a\Big{[}6(\textbf{n}_{bE}\cdot\textbf{x}_{A})(\textbf{n}_{bE}\cdot\textbf{v}_{B})+6(\textbf{n}_{bE}\cdot\textbf{x}_{B})(\textbf{n}_{bE}\cdot\textbf{v}_{A})
 \nonumber\\
  \!\!\!\!\!\!\!\!\!\!\!\!\!\!\!\!\!\!\!\!\!\!\!\!\!\!\! +4(\textbf{n}_{bE}\cdot\textbf{R}_{AB})(\textbf{n}_{bE}\cdot\textbf{v}_{AB})-\sqrt{2GMa}e\Big{(}\sin\Big{(}E_{A}+\frac{\pi}{4}\Big{)}+\sin\Big{(}E_{B}+\frac{\pi}{4}\Big{)}\Big{)}\Big{]}\\
  \!\!\!\!\!\!\!\!\!\!\!\!\!\!\!\!\!\!\!\!\!\!\!\!\!\!\!-\sqrt{\frac{3GM}{4a}}e\sin E_{AB}\Big{[}6(\textbf{n}_{bE}\cdot\textbf{x}_{B})(\textbf{n}_{bE}\cdot\textbf{x}_{A})
  +2(\textbf{n}_{bE}\cdot\textbf{R}_{AB})^{2}-5a^{2}\Big{]}\Big{\}}+O(e^{2},r^{-4}_{bE}).\nonumber
\end{eqnarray}
where the $E_{A/B}$ is the eccentric anomaly, and $E_{AB}=(1/2)(E_{A}+E_{B})$. Eqs.(\ref{dtidefull1ab}) and(\ref{dttftidab}) can be used to calculate tidal effects on the propagation of signal, however Eq.(\ref{dtidefull1ab}) is more precise in the case of high orbit and Eq.(\ref{dttftidab}) is more convenient to estimate tidal contributions for low orbit.

\begin{table}[!t]
\flushright
\caption{\label{tab:1}The Effects due to the Earth and other solar system bodies in one laser arm for TianQin mission. The semi-major axis of satellites orbit is 1$\times 10^{5}$km and the orbital eccentricity is 0.01. }
\newcommand{\tabincell}[2]{\begin{tabular}{@{}#1@{}}#2\end{tabular}}
\begin{tabular}{lccc}
\hline
\tabincell{l}{Effect} &Contribution \\
\hline
Frequency shift due to the Earth &$4\times10^{-17}$\\
Frequency shift due to the astronomical
tidal potentials  &$\sim10^{-19}$\\
Frequency shift due to the Earth's $J_{n}$  &$\sim10^{-22}$\\
Frequency shift due to the Earth's spin &$\sim10^{-23}$\\

\hline
\end{tabular}
\end{table}

In order to calculate the frequency shift due to the gravitational delay, the derivatives of the terms $\Delta_{M_{E}}$, $\Delta_{J_{n}}$ and $\Delta_{s}$ should be determined. From the expressions of gravitational delay in Ref.\cite{shapiro,ad12,ad14}, we can obtain direct the derivatives. Inserting these derivatives into Eq.(\ref{spacefreq}), we estimate frequency shift due to gravitational delay for TianQin mission (Table \ref{tab:1}). For example, the first term in Table \ref{tab:1} is the contribution of mass monopole, which can be expressed as
\begin{eqnarray}\label{deltmtotal}
  \Big{(}\frac{\delta\nu}{\nu}\Big{)}_{\Delta_{M_{E}}}
  =\frac{4\sqrt{3}GM_{E}e}{ac^{3}}\sqrt{\frac{GM_{E}}{a}}(\sin E_{AB}+\sin E_{A}+\sin E_{B}).
\end{eqnarray}
The orbit-dependent equation implies that this effect may be amplified or inhibited through choosing suitable orbital parameter.

In the space missions like TianQin or GPS, the recording time of signal is in one of satellites. Therefore, it is convenient to express the time transfer function(\ref{ttfd2}) at the time of emission $t_{A}$ or at the time of reception $t_{B}$. We introduce the case of the reception time $t_{B}$, the case of emission time $t_{A}$ can be analyzed with the same way. We introduce an instantaneous coordinate distance $D_{AB}=|\textbf{D}_{AB}|$ with $\textbf{D}_{AB}=\textbf{x}_{B}(t_{A})-\textbf{x}_{A}(t_{A})$. The Taylor expansion of $\textbf{R}_{AB}$ can be written as
\begin{equation}\label{ttftl}
\textbf{R}_{AB}\equiv\textbf{x}_{B}(t_{B})-\textbf{x}_{A}(t_{A})
=\textbf{D}_{AB}+\textbf{v}_{A}(t_{A})T_{AB}+\frac{1}{2}\textbf{a}_{A}(t_{A})T^{2}_{AB}+\frac{1}{6}\textbf{b}_{A}(t_{A})T^{3}_{AB}+O(c^{-4}),
\end{equation}
where $\textbf{v}_{A}$ is the velocity of $A$, $\textbf{a}_{A}$ is the acceleration of $A$ and $\textbf{b}_{A}=d\textbf{a}_{A}/dt$. By an iterative process, $R_{AB}$ can be rewritten as
\begin{eqnarray}\label{ttfrdab}
   \!\!\!\!\!\!\!\!\!\!\!\!\!\!\!\!\!\!\!\!\!\!\!\!\!\!\!\!\!\!\!\!\!\!\!\!R_{AB}=D_{AB}+\frac{\textbf{D}_{AB}\cdot\textbf{v}_{A}}{c}
  +\frac{D_{AB}}{2c^{2}}\Big{[}\textbf{D}_{AB}\cdot\textbf{a}_{A}+\textbf{v}_{A}^{2}+\frac{(\textbf{D}_{AB}\cdot\textbf{v}_{A})^{2}}{D^{2}_{AB}}\Big{]}\\
    \!\!\!\!\!\!\!\!\!\!\!\!\!\!\!\!\!\!\!\!\!\!\!\!\!\!\!\!\!\!\!\!\!\!\!\!+\frac{1}{c^{3}}\Big{(}\frac{1}{6}D_{AB}^{2}\textbf{D}_{AB}\cdot\textbf{b}_{A}+\frac{1}{2}D_{AB}^{2}\textbf{a}_{A}\cdot\textbf{v}_{A}
  +\textbf{D}_{AB}\cdot\textbf{v}_{A}\textbf{v}^{2}_{A}+\textbf{D}_{AB}\cdot\textbf{v}_{A}\textbf{D}_{AB}\cdot\textbf{a}_{A}\Big{)}+O(c^{-4}),\nonumber
\end{eqnarray}
where all the quantities measured at reception instant time $t_{A}$. The second, third and fourth terms of this equation represent the Sagnac terms of order $c^{-2}$, $c^{-3}$, and $c^{-4}$, respectively. For TianQin space mission, the second term can be reexpressed as
\begin{eqnarray}\label{sagnacre2}
  \frac{\textbf{D}_{AB}\cdot\textbf{v}_{A}}{c}=\frac{\sqrt{3GMa}}{2c}\Big{[}1+\frac{\Delta a}{2a}+(e+\Delta e)\cos E_{A}\Big{]},
\end{eqnarray}
where $\Delta a$, $\Delta e$ are given by Eqs.(\ref{daaver}) and (\ref{dedp}), respectively.  This term is about 5.8$\times10^{2}$m. Since LISA requires that the difference in length of the two arms is known to 200m \cite{wave2}, TianQin has the similar requirement for the arm length. The Sagnac term should be carefully treated in TianQin.

The time transfer function $T_{AB}$ in the vicinity of the Earth can be expressed as
\begin{eqnarray}\label{sp3}
  T_{AB}(\textbf{D}_{AB})=\frac{R_{AB}(\textbf{D}_{AB})}{c}+\frac{\Delta_{r}(\textbf{D}_{AB})}{c}.
\end{eqnarray}
From the Eqs.(\ref{ttfd2}), (\ref{sp01}) and (\ref{sp001}), the frequency shift can be determined at the order of $c^{-4}$. Using Eq.(\ref{ttfrdab}), we obtain
\begin{equation}\label{ttfsagnac}
  \frac{dR_{AB}}{dct_{A}}=\frac{\textbf{n}_{AB}\cdot\textbf{v}_{AB}}{c}+\frac{1}{c^{2}}(\textbf{v}_{AB}\cdot\textbf{v}_{A}+\textbf{D}_{AB}\cdot\textbf{a}_{A})
  +\delta _{sag3}+\delta_{sag4}+O(c^{-5}),
\end{equation}
where $\textbf{v}_{AB}=d\textbf{D}_{AB}/dt_{A}$. For the GPS with the orbital radius $2.66\times10^{7}$m, the upper bound of first term is $1.5\times10^{-5}$ and the upper bound of the second term is $3.2\times10^{-10}$. The third term represents the Sagnac contribution of the order $c^{-3}$, which is given by
\begin{eqnarray}\label{sagnac3er}
   \!\!\!\!\!\!\!\!\!\!\!\!\!\!\!\!\!\!\delta _{sag3}=\frac{\textbf{n}_{AB}\cdot\textbf{v}_{AB}}{2c^{3}}[\textbf{D}_{AB}\cdot\textbf{a}_{A}+\textbf{v}_{A}^{2}-(\textbf{n}_{AB}\cdot\textbf{v}_{A})^{2}]\\
   \!\!\!\!\!\!\!\!\!\!\!\!\!\!\!\!\!\!+\frac{D_{AB}}{2c^{3}}\Big{[}\textbf{D}_{AB}\cdot\textbf{b}_{A}+\textbf{v}_{AB}\cdot\textbf{a}_{A}+2\textbf{v}_{A}\cdot\textbf{a}_{A}
  +2\frac{(\textbf{D}_{AB}\cdot\textbf{v}_{A})(\textbf{D}_{AB}\cdot\textbf{a}_{A}+\textbf{v}_{AB}\cdot\textbf{v}_{A})}{D^{2}_{AB}}\Big{]}.\nonumber
\end{eqnarray}
The contribution of this term may reach the level of $10^{-15}$. The fourth term represents the Sagnac contribution of the order $c^{-4}$, whose contribution is negligible.

Assuming that $A$ is the ground station and $B$ is the satellite, $\boldsymbol{\omega}_{E}$ is the angular-rotation-rate vector of the Earth. The leading contribution of Sagnac term becomes
\begin{eqnarray}\label{ess331}
   \Delta_{sag}=\frac{\textbf{x}_{B}\cdot(\boldsymbol{\omega}_{E}\times\textbf{x}_{A})}{c}=\frac{2\omega_{E} S}{c},
\end{eqnarray}
where $S$ is the projection onto the equatorial plane of the triangle $\triangle EOS$ area. The triangle $\triangle EOS$ consists of the center of the Earth, the observer and the satellite.
When GPS is on the equatorial plane and the terrestrial station is on the equator, the Sagnac delay reaches several 10m.

\subsection{The frequency shift for the GPS-like space missions}\label{V}
The method of Two-Way satellite has been applied to make intercontinental clock and frequency comparisons with a better accuracy. For the applications of the GPS communications and space missions like ACES mission \cite{aces}, it is necessary to implement frequency comparisons between the clocks on satellites and clocks on the ground. The influence of relativistic effects is a crucial part and the rigorous calculation is requisite.
For example, the difference in rate between GPS satellite and ground clocks is 46$\mu$s per day from the gravitational refshift and -7$\mu$s per day from time dilation.
In this section, we pay attention to relativistic effects for clocks on the perturbed Kepelerian orbit. In the GCRS framework, we suppose a homogeneous medium with the effective refractive index $n$. Reanalyzing Eqs.(\ref{sp01})-(\ref{k0ab}), the one-way frequency shift may be written as
\begin{equation}\label{gsfretrans1}
 \frac{\nu_{B}}{\nu_{A}}=\frac{(g_{00}+2g_{0i}p^{i}+g_{ij}p^{i}p^{j})^{1/2}_{A}}{(g_{00}+2g_{0i}p^{i}+g_{ij}p^{i}p^{j})^{1/2}_{B}}
  \frac{1-N^{i}_{AB}p^{i}_{B}-p^{i}_{B}\frac{\partial\Delta_{r}}{\partial x^{i}_{B}}-\frac{n\partial\Delta_{r}}{c\partial t_{B}}-\frac{\partial n}{c\partial t_{B}}R_{AB}}{1-N^{i}_{AB}p^{i}_{A}+p^{i}_{A}\frac{\partial\Delta_{r}}{\partial x^{i}_{A}}}.
\end{equation}
Then, we consider the case that effective refractive index is not constant. We consider the effective refractive index evolving with time and take its heterogeneity on the light propagation into account. For a reasonable approximation, the frequency shift may be written as
\begin{eqnarray}\label{gsfretrans2}
   \frac{\nu_{B}}{\nu_{A}}=\Big{(}\frac{u^{0}_{B}}{u^{0}_{A}}\frac{q_{B}}{q_{A}}\Big{)}_{vac}-\frac{1}{c}\int\frac{\partial n(l,t)}{\partial t}dl
\end{eqnarray}
with $l\in[0,R_{AB}]$. The second term is the atmospheric frequency shift, which is dependent on the light propagation paths in the atmosphere. The atmospheric frequency shift need the specific model to study, however it is beyond the scope in this section.

The first term of Eq.(\ref{gsfretrans2}) is the vacuum frequency shift, which is equivalent to Eq.(\ref{spacefre}). We can estimate Earth gravitational field's and tidal potential's influences on time transfer and frequency shift. Since Sec.\ref{IV} has discussed the frequency shift due to gravitational delay in the term $q_{B}/q_{A}$, we will focus on the relativistic tidal effects of term $u^{0}_{B}/u^{0}_{A}$ in this section. As mentioned in Introduction, for the on-board clock, tidal effects are divided into two parts: tidal potential itself (we name this part tidal gravitational redshift in following text) and orbital changes due to tidal force.

Firstly, we consider the tidal effect due to orbital changes. The tidal forces can change the velocity and position of satellites, which in turn leads to perturbing effects in Earth's gravitational redshift $(\delta f/f)_{Eper}$ and second Doppler effect $(\delta v^{2}/c^{2})_{per}$. For a GPS satellite with orbital radius 2.66$\times10^{7}$m, it gives that the tidal acceleration due to Moon is the level of $6.9\times10^{-6}$m/s$^{2}$ whereas for Sun it is $3.2\times 10^{-6}$m/s$^{2}$.  For an interval of 1000s, it is direct to deduce the levels of perturbing gravitational redshift of Earth $2\times10^{-17}$ and $1\times10^{-17}$ for Moon's tidal acceleration and Sun's tidal acceleration, respectively. For an interval of several hours, the perturbations would reach several $10^{-16}$. It demonstrates that we should carefully treat the orbital changes due to tidal forces. We treat it as a perturbed Kepler problem.
The perturbed Kepler problem is presented with the $\emph{method of osculating orbital}$ $\emph{elements}$ in \ref{appa}. We assume that unperturbed Kepler orbit is
$r= p(1+e\cos{f})^{-1},$
where $e$ is the eccentricity, and $p$ is orbit's semi-latus rectum. For the perturbed Kepler problem, we let the orbital elements become functions of time. The perturbed Kepler orbit may be expressed as the form
\begin{eqnarray}\label{unpkr}
  r(t)=\frac{p(t)}{1+e(t)\cos{f(t)}}.
\end{eqnarray}
The $\emph{method of osculating orbital elements}$ means that the orbit is regarded as a Keplerian orbit with orbital elements $p(t_{1}), e(t_{1})...$ at time $t_{1}$ only; orbit is a still Keplerian orbit at another time $t_{2}$, but the orbital elements have evolved to new values $p(t_{2}), e(t_{2})...$ And at a given time, the expressions for velocity and radial position of Keplerian orbit are sufficient with the orbital elements of that time. The secular variation of orbital elements between time $t_{1}$ and time $t_{2}$ can be calculated by integrating Eqs.(\ref{deldt})-(\ref{deldtit}) from $t_{1}$ to $t_{2}$. A useful expression for velocity and position is
\begin{eqnarray}\label{vpt}
  v^{2}(t)=GM_{E}\Big{(}\frac{2}{r(t)}-\frac{1}{a(t)} \Big{)}.
\end{eqnarray}

The second-order Doppler effect and gravitational redshift depend on the velocity and position, respectively. The tidal perturbing forces change position and velocity of the satellite, which leads to a perturbing contributions in the second-order Doppler effect and a perturbing contribution in the Earth's gravitational redshift. For perturbed Kepler orbit, the gravitational redshift due to Earth is $GM_{E}/c^{2}r(t)$. We consider the perturbing force is decomposed as
\begin{eqnarray}\label{perf1}
  \textbf{f}=\mathcal{A}\textbf{n}+\mathcal{B}\textbf{n}_{\bot}+\mathcal{C}\textbf{e}_{z},
\end{eqnarray}
where $\textbf{n}$ is the unit vector that points from Earth to satellite, $\textbf{e}_{z}$ is the unit vector orthogonal to orbital plane, and $\textbf{n}_{\bot}$ is the unit vector that is orthogonal to $\textbf{n}$ with relationship $\textbf{n}\times\textbf{n}_{\bot}=\textbf{e}_{z}$.
The perturbation in Earth's gravitational redshift from perturbing forces is
\begin{eqnarray}\label{eredper}
  \Big{(}\frac{\delta f}{f}\Big{)}_{Eper}=\frac{GM_{E}}{c^{2}r}\Big{(}\frac{\Delta e(t)\cos{f}-e\Delta f(t)\sin{f}}{1+e\cos{f}}-\frac{\Delta p(t)}{p}\Big{)},
\end{eqnarray}
where $\Delta e(t)$, $\Delta p(t)$ and $\Delta f(t)$ are given by
\begin{eqnarray}\label{dedp}
  \Delta e(t)=\frac{\Delta t}{P}\int^{P}_{0}\sqrt{\frac{p}{GM_{E}}}\Big{[}\sin{f}\mathcal{A}+\frac{2\cos{f}+e(1+\cos^{2}f)}{1+e\cos{f}}\mathcal{B}\Big{]}dt,\\
  \Delta p(t)=\frac{\Delta t}{P}\int^{P}_{0}2\sqrt{\frac{p^{3}}{GM_{E}}}\frac{\mathcal{B}}{1+e\cos{f}}dt,
\end{eqnarray}
\begin{equation}\label{dedfintef}
 \Delta f(t)=\frac{\Delta t}{P}\int^{P}_{0}\sqrt{\frac{Gm}{p^{3}}}(1+e\cos{f})^{2}+\frac{1}{e}\sqrt{\frac{p}{Gm}}\Big{[}\cos{f}\mathcal{A}-\frac{2+e\cos{f}}{1+e\cos{f}}\sin{f}\mathcal{B} \Big{]}dt,
\end{equation}
where $\Delta t\gg P$. When $\Delta t\sim P$, the variation of orbital elements is given by integrating Eqs.(\ref{deldt})-(\ref{deldtit}) from time $t$ to time $t+\Delta t$. For the purpose of numerical analysis or more precise expression, these three equations should be given by the iteratively integrating Eqs.(\ref{deldt})-(\ref{deldtit}).

With the help of Eq. (\ref{vpt}), we can also calculate the perturbation in the second Doppler effect due to perturbing force
\begin{eqnarray}\label{v2per}
\Big{(}\frac{\delta v^{2}}{c^{2}} \Big{)}_{per}=\frac{2GM_{E}}{c^{2}r}\Big{(}\frac{\Delta e(t)\cos{f}-e\sin{f}\Delta f(t)}{1+e\cos{f}}-\frac{\Delta p(t)}{p}+\frac{r\Delta a(t)}{2a^{2}}\Big{)},
\end{eqnarray}
where $\Delta a(t)$ is given by
\begin{eqnarray}\label{daaver}
  \Delta a(t)=\frac{\Delta t}{P}\int^{P}_{0}2\sqrt{\frac{a^{3}}{GM_{E}}}\frac{e\sin{f}\mathcal{A}+(1+e\sin{f})\mathcal{B}}{\sqrt{1-e^{2}}}dt.
\end{eqnarray}
For the purpose of more precise expression, it also should be given by the iteratively integrating Eqs.(\ref{deldt})-(\ref{deldtit}).
It is sufficient to calculate relativistic orbital effects with this formalism, such as, gravitational redshift and the second Doppler effect. When the perturbing force is tidal force from the Moon, the parameters $\mathcal{A}$, $\mathcal{B}$ and $\mathcal{C}$ are given by Eqs. (\ref{tkpa})-(\ref{tkpc}).

As an example, we estimate the contribution of the Moon and assume that the orbital plane of satellite coincides with that of Moon. It means that the inclination of the orbital plane to reference plane $i$ is zero. The components of Moon's tidal force become the simpler forms. With the variations of orbital elements (\ref{elementd1})-(\ref{elementd2}), the perturbing contributions of Earth's gravitational redshift and second Doppler effect due to Moon's tidal force are
\begin{eqnarray}\label{simperedper}
  \!\!\!\!\!\!\!\!\!\!\!\!\!\!\!\!\!\!\!\!\!\!\!\!\!\!\!\Big{(}\frac{\delta f}{f}\Big{)}_{Eper,m}=\frac{15\pi GM_{m}}{c^{2}r^{3}_{m}}\frac{a^{2}e}{\sqrt{1-e^{2}}}
  \Big{[}\frac{\sin{2\Psi}\cos{f}}{2}+\frac{e\sin{2\Psi}(1+e\cos{f})}{1-e^{2}}\Big{]}\\
  \!\!\!\!\!\!\!\!\!\!\!\!\!\!\!\!\!\!\!\!\!\!\!\!\!\!\!=2.3\times10^{-29}\frac{a^{2}e}{\sqrt[3]{1-e^{2}}}
   [\sin{2\Psi}\cos{f}+2e\sin{2\Psi}+e^{2}\sin{2\Psi}\cos{f}]\nonumber
\end{eqnarray}
and
\begin{eqnarray}\label{simpv2per}
    \!\!\!\!\!\!\!\!\!\!\!\!\!\!\!\!\!\!\!\!\!\!\!\!\!\!\!\Big{(}\frac{\delta v^{2}}{c^{2}} \Big{)}_{per,m}=\frac{15\pi GM_{m}}{c^{2}r^{3}_{m}}\frac{a^{2}e}{\sqrt{1-e^{2}}}
  \Big{[}\sin{2\Psi}\cos{f}+\frac{2e\sin{2\Psi}(1+e\cos{f})}{1-e^{2}}\Big{]}\\
  \!\!\!\!\!\!\!\!\!\!\!\!\!\!\!\!\!\!\!\!\!\!\!\!\!\!\!=4.6\times10^{-29}\frac{a^{2}e}{\sqrt[3]{1-e^{2}}}
   [\sin{2\Psi}\cos{f}+2e\sin{2\Psi}+e^{2}\sin{2\Psi}\cos{f}]\nonumber
\end{eqnarray}
after a complete satellite orbit, where $\Psi=(\omega-F+\Omega)$. Eq.(\ref{simperedper}) gives the perturbing effect of Earth's gravitational redshift due to the fact that Moon's tidal force changes satellite's position after a complete orbital period. Eq.(\ref{simpv2per}) gives the perturbing effect of the second Doppler effect due to the fact that Moon's tidal force changes satellite's velocity after a complete orbital period.
For the GPS satellite orbit with the eccentricity $e=0.01$, the perturbing contributions of Earth's gravitational redshift and second Doppler effect due to Moon's tidal force can reach the level of $1.6\times10^{-16}$ and $3.2\times 10^{-16}$, respectively. The perturbations of Earth's gravitational redshift and second Doppler effect after a complete satellite orbit are smaller than that after only several hours since perturbations may partly cancel for the different positions in the orbit. These perturbations may reach the level of several $10^{-16}$ even $1\times 10^{-15}$  for several hours. As mentioned in Ref.\cite{002}, tidal forces may lead to 50-150m position error for three hours of GPS satellite moving time, which can change the gravitational redshift of Earth with $(3\sim9)\times 10^{-16}$. These formulae may be applied to calculate the perturbation due to Sun through replacing subscript $m$ by $s$. The subscript $s$ represents quantities of the Sun.

Secondly, we consider the tidal effect due to tidal potential itself (or the tidal gravitational redshift). Combing Eq.(\ref{tidee}) and Legendre polynomials, it can be written as
\begin{equation}\label{ggred4}
  \delta_{tidal}=\sum_{b\neq E}\frac{GM_{b}a^{2}}{2c^{2}r^{3}_{bE}}(3(\textbf{n}_{bE}\cdot\textbf{n})^{2}-1)(1+2e\cos E+e^{2}\cos^{2} E)+O(r^{-4}_{bE}).
\end{equation}
The term $\textbf{n}_{bE}\cdot\textbf{n}$ may be expressed as
\begin{eqnarray}\label{ggred5}
  \textbf{n}_{bE}\cdot\textbf{n}=\cos z_{b} \cos z_{s}+ \sin z_{b} \sin z_{s} \cos H_{bs},
\end{eqnarray}
where $z_{b}$  is the zenith angle of the body $b$, $z_{s}$ is the zenith angle of the satellite, and $H_{bs}$ is a angle between body $b$ and satellite with the same period to satellite's orbital period, whose definition is similar to local hour angle by the observer. When we consider the Moon, $z_{b}$ may be given as Eq.(\ref{8}) and $\cos z_{b}, \sin z_{b}$ may be determined by lunar calendar. Similarly, $\cos z_{s}, \sin z_{s}$ may be given by satellite ephemeris. It is sufficient to use $\cos z_{s}=\sin \varphi \sin \delta_{s}+\cos \varphi \cos \delta_{s} \cos H_{s}$ for the satellite, where the $\varphi$ is the latitude of observer, $H_{s}$ and $\delta_{s}$ is the local hour angle and declination of the satellite, respectively. Combining Eqs.(\ref{5t}) and(\ref{ggred5}),the tidal gravitational redshift on clock comparisons between ground and satellites may be given by
\begin{eqnarray}\label{ggred6}
 \!\!\!\!\!\!\!\!\!\!\!\! \Big{(}\frac{\delta f}{f}\Big{)}_{tidal}=\sum_{b\neq E}
  \frac{2D_{b}}{c^{2}}\Big{\{}\Big{(}\frac{a}{R_{E}}\Big{)}^{2}\Big{(}\frac{c_{b}}{r_{bE}}\Big{)}^{3}\Big{(}(\textbf{n}_{bE}\cdot\textbf{n})^{2}-\frac{1}{3}\Big{)}(1+2e\cos E+e^{2}\cos^{2} E)\nonumber\\
  -(1-h_{2}+k_{2})\Big{(}\frac{r_{g}}{R_{E}}\Big{)}^{2}\Big{(}\frac{c_{b}}{r_{bE}}\Big{)}^{3}\Big{(}\cos^{2}z_{b}-\frac{1}{3}\Big{)}\Big{\}},
\end{eqnarray}
where $r_{g}=|\textbf{x}_{g}|$, $\textbf{x}_{g}$ is the position of ground clock, $D_{b}$ is the Doodson constant of body $b$, $z_{b}$ is the zenith angle of the body $b$, and $c_{b}$, $r_{b}$ are the average distance and practical distance of the body $b$ to the center of Earth mass, respectively.

\section{Conclusion}\label{VI}
In this work we present a relativistic procedure to rigorously deduce the frequency shift up to the order $c^{-4}$ for clock-comparison experiments in the vicinity of the Earth and focus on formulating relativistic tidal effects in frequency shift.

For clock-comparison experiments linked with optical fibres on the ground, we focus on the relativistic tidal effects due to the tidal potential and tidal response of solid Earth. When the target accuracy is the level of $10^{-19}$ , Love numbers are sufficient to describe the influence of tidal response of solid Earth, and the influence of the optical fibres is obtained up to the order of $c^{-4}$. In order to demonstrate properties of tidal effect, we study the influences of the tidal potentials in the celestial coordinate system, which explicitly shows the diurnal, semidiurnal and one-third-diurnal variations. Moreover, after considering the tidal response of the solid Earth, we formulate mathematical connection between the local gravity tides from the gravimeters and tidal effects from the clock-comparison experiments on the ground (see \ref{apptidal}). It demonstrates that the signal of tidal influences on clocks coincides with that of corresponding-position gravity tides, which further provides us an approach with local tides data to eliminate the tidal influences in frequency and clock comparisons. For the current accuracy of clock comparison in laboratories, Eq.(\ref{16a}) is sufficient to calculate tidal effects, while the third-order tidal potential should been take into account for future high-precision experiments. To demonstrate our results, we simulate the relativistic tidal effects. It demonstrates that the contributions due to the total second-order tidal potentials can reach $1\times10^{-17}$ for frequency comparisons with the distances $\sim10^{3}$km, and contributions due to the third-order tidal potential can reach $2\times10^{-19}$. This means the second-order tidal effects is measurable in laboratories for clock comparisons at the distance of several 100km.

For clock-comparison experiments in the deep space missions, we use the TTF formalism to study frequency shift and time transfer in the gravitational field of a tidal, axisymmetric, rotating Earth, and give corresponding estimates for TianQin mission. The tidal gravitational delay and corresponding frequency shift are derived, which may respectively reach $10^{-6}$m level in distances and $10^{-19}$ in frequency shift for each laser arm of TianQin mission. To well model general relativistic observables, the frequency shift due to gravitational delay is given by the orbit-parameter-dependence form. For TianQin mission, Earth's influences are partly listed in Table \ref{tab:1}. Furthermore, we give a rough estimate for the Sagnac effect and find the third Sagnac term needs to be considered for some modern experiments.

For the GPS-like space missions, we develop a method of perturbed Kepler orbit to determine relativistic frequency shift for clock comparison. Since the perturbing forces would change the position and velocity of satellite, it is insufficient to calculate the relativistic effects with the Keplerian orbit. Comparing with the conventional method of Keplerian orbit, the perturbed Kepler-orbit method can model more precise  relativistic observables of tidal effects for on-board clocks.
To account for tidal effects, one should notice that there is an first effect from the tidal potential itself, and then effects of similar orders of magnitude from changes in position and velocity due to the tidal perturbing forces.  The former one is parameterized by Eq.(\ref{ggred6}), and the latter one would lead to the perturbations in the Earth's gravitational redshift and the second-order Doppler effect, which is given by Eqs.(\ref{eredper}) and (\ref{v2per}) through the method of perturbed Kepler orbit. Finally, with the development of space technology, the precise relativistic modeling of observables should include more perturbing forces.

\section{Acknowledgment}
This work is supported by the National Natural Science Foundation of China(Grant Nos. 91636221 and 11805074), and the Post-doctoral Science Foundation of China (Grant Nos. 2017M620308 and 2018T110750).

\appendix
\section{Time transfer function}\label{timetf}
Let us consider two observers $O_{A}$ and $O_{B}$ located at points $\textbf{x}_{A}$ and $\textbf{x}_{B}$, respectively. We suppose that a light signal is emitted by the observer $O_{A}$ and is received by the observer $O_{B}$. In other words, it supposes that $x_{A}=(ct_{A},\textbf{x}_{A})$ and $x_{B}=(ct_{B},\textbf{x}_{B})$ are two event points of the spacetime which are connected by a unique light ray. The light propagation in the gravitational filed can be studied by null geodesic Eqs.\cite{null1,null2,null3,null4} or time transfer functions \cite{ad2,adt1,adt2,adt3}. Some solutions have been proposed in the post-Minkowskian and post-Newtonian approximations in the Solar System. The difference $t_{B}-t_{A}$ is the coordinate travel time of a light signal connecting event points $x_{A}$ and $x_{B}$. This quantity may be written as a time transfer function:
\begin{eqnarray}\label{ttfd1}
  t_{B}-t_{A}=T_{r}(\textbf{x}_{A},t_{B},\textbf{x}_{B})=T_{e}(t_{A},\textbf{x}_{A},\textbf{x}_{B}),
\end{eqnarray}
where $T_{r}$ and $T_{e}$ are the reception time transfer function and emission time transfer function, respectively. In the following, we consider only the case of the reception TTF and the discussion can be done with the same way used in the case of the emission TTF (from the experimental point of view, the observables are recorded at the time of reception $t_{B}$). In the weak field approximation, the reception time transfer function may be written as follows \cite{ad3,ad4}:
\begin{eqnarray}\label{ttfd2}
T_{r}(\textbf{x}_{A},t_{B},\textbf{x}_{B})=\frac{R_{AB}}{c}+\frac{\Delta_{r}(\textbf{x}_{A},t_{B},\textbf{x}_{B})}{c},
\end{eqnarray}
where $R_{AB}=|\textbf{x}_{B}-\textbf{x}_{A}|$, and $\Delta_{r}(\textbf{x}_{A},t_{B},\textbf{x}_{B})$ is the so-called "delay function."

The time transfer function can be determined by the iterative method used in the Ref.\cite{ad3}. It satisfies the Hamilton-Jacobi-like equation:
\begin{equation}\label{hamilton}
  g^{00}(x^{0}_{B}-cT_{r},\textbf{x}_{A})+2cg^{0i}(x^{0}_{B}-cT_{r},\textbf{x}_{A})\frac{\partial T_{r}}{\partial x^{i}_{A}}
  +c^{2}g^{ij}(x^{0}_{B}-cT_{r},\textbf{x}_{A})\frac{\partial T_{r}}{\partial x^{i}_{A}}\frac{\partial T_{r}}{\partial x^{j}_{A}}=0.
\end{equation}
In the Solar System, it is sufficient to suppose that metric and delay function are represented at any point $x$ by a series in ascending powers of the Newtonian gravitational constant $G$\cite{ad3}:
\begin{eqnarray}\label{conmetric1}
  g^{\mu\nu}(x,G)=\eta^{\mu\nu}+\sum^{\infty}_{n=1}G^{n}g_{(n)}^{\mu\nu}(x),
\end{eqnarray}
and
\begin{eqnarray}\label{conmetric2}
  \Delta_{r}(\textbf{x},t_{B},\textbf{x}_{B},G)=\sum^{\infty}_{n=1}G^{n}\Delta^{(n)}_{r}(\textbf{x},t_{B},\textbf{x}_{B}),
\end{eqnarray}
where $\eta^{\mu\nu}=$diag(1,-1,-1,-1). In the post-Minkowskian approximation, the Eqs.(\ref{hamilton})-(\ref{conmetric2}) allow to determine delay function by the integral of the metric components over a straight line between the emitter and the receiver of a light signal. In terms of experiment's applications, the case $n\geq 3$ of delay function is completely negligible since $n=3$ corresponds to the terms $O(c^{-6})$ . Therefore, delay function may be rewritten as $\Delta_{r}=G\Delta^{(1)}_{r}+G^{2}\Delta^{(2)}_{r}+O(G^{3}),$, which is given by Ref.\cite{ad3}:
\begin{eqnarray}\label{generalfun}
 \!\!\!\!\!\! \Delta^{(1)}_{r}=\frac{R_{AB}}{2}\int^{1}_{0}(g_{(1)}^{00}-2N^{i}_{AB}g_{(1)}^{0i}+N^{i}_{AB}N^{j}_{AB}g_{(1)}^{ij})_{x(\lambda)}d\lambda\nonumber\\
 \!\!\!\!\!\! \Delta^{(2)}_{r}=\frac{R_{AB}}{2}\int^{1}_{0}\Big{[}(g^{00}_{(2)}-2N^{i}_{AB}g^{0i}_{(2)}+N^{i}_{AB}N^{j}_{AB}g^{ij}_{(2)})_{x(\lambda)}   \\
+2(g^{0i}_{(1)}-N^{j}_{AB}g^{ij}_{(1)})_{x(\lambda)}\frac{\partial\Delta_{r}^{(1)}}{\partial x^{i}}(x(\lambda),t_{B},\textbf{x}_{B})
+\eta^{ij}\Big{(}\frac{\partial\Delta_{r}^{(1)}}{\partial x^{i}}\frac{\partial\Delta_{r}^{(1)}}{\partial x^{j}}\Big{)}_{(x(\lambda),t_{B},\textbf{x}_{B})}\Big{]}d \lambda, \nonumber
\end{eqnarray}
where $N_{AB}^{i}=(x_{B}^{i}-x_{A}^{i})/R_{AB}$, and the integral is taken along the segment of a line of ends $\textbf{x}_{A}$ and $\textbf{x}_{B}$ described by the parametric equations
\begin{eqnarray}\label{parameter0}
  x^{0}(\lambda)=ct_{B}-\lambda R_{AB},\,\,\,\,\,\,\,\,  \textbf{x}(\lambda)=\textbf{x}_{B}-\lambda \textbf{R}_{AB},  (0\leq\lambda\leq1.)
\end{eqnarray}
The $n$th term of delay function may be deduced by first $n-1$ terms and their derivatives. Eq.(\ref{generalfun}) is sufficient for experiment in the vicinity of Earth.

\section{The tidal response of solid Earth}\label{apptidal}
It is well known that motions induced in the solid Earth by tidal forces are the Earth tides. Due to the elasticity of Earth, tidal forces produce deformations in the Earth and ocean. The density of the solid Earth also is affected under the influence of the tidal forces. Since 1 mm change in height leads to a frequency shift $\sim1\times10^{-19}$, the tidal response of the solid Earth should be considered for the time and frequency comparisons. The tidal response of the solid Earth is based on an Earth which is spherical, non-rotating, elasticity, isotropic and oceanless. Firstly, we discuss Moon's tidal potential, then we can similarly analyze that of Sun. In GCRS framework, Moon's gravitational potential can be expanded as the form of $U_{E}$. Then, tidal potential of Moon may be written as
\begin{eqnarray}\label{6}
  u^{tidal}_{m}(\textbf{x})=\sum^{\infty}_{n=2}u^{tidal}_{nm}(\textbf{x})=\frac{GM_{m}}{r_{m}}\sum^{\infty}_{n=2}\Big{(}\frac{r}{r_{m}}\Big{)}^{n}P_{n}(\cos z_{m}),
\end{eqnarray}
where $M_{m}$ is the mass of Moon, $r_{m}$ is the practical distance from Moon to Earth's center of mass, $z_{m}$ is the zenith angle of the Moon, $P_{n}(\cos z_{m})$ are the Legendre polynomials. It is clear that the Moon's tidal potential depends on distance and position of Moon. The effects of tidal potential are variable with the position of celestial bodies.

The Moon's tidal force results Earth's gravitational potential a small change on the terrestrial surface. This variation primarily comes from two parts: (1) the deformation of Earth (terrestrial height's fluctuation in the influence of Moon's tidal force); (2) the redistribution of Earth mass. Each of parts possesses its own expression in term of Love numbers. Love numbers describe Earth's response to Moon and Sun and are just the function of $r$. On terrestrial surface, the part (1) is expressed as a variation of height, which is given by
\begin{eqnarray}\label{10}
  \delta r_{m}=\sum^{\infty}_{n=2}h_{n}\frac{u^{tidal}_{nm}(\textbf{x})}{g_{0}},
\end{eqnarray}
where $h_{n}$ is the $n$th-order Love number on terrestrial surface, which describes fluctuation of height due to the Moon's tidal force, and $g_{0}$ is average gravity on the terrestrial surface. According to this equation, height has a periodic change in that frequency with the same phase as the periodic variation of the Moon's tidal potential. Magnitude of $\delta r_{m}$ is $10^{-1}$m, which is in the measurable range of geodesy. This simple expression allows us to study contribution due to change of height. Considering the part (1), we can express Earth's gravitational potential on the terrestrial surface as a sum of a constant potential $U_{E}(\textbf{x})$ and a variable potential $\delta_{m} U_{E}(\textbf{x})$
\begin{eqnarray}\label{11}
U_{E}(\textbf{x}+\delta\textbf{x}_{m})=U_{E}(\textbf{x})+\delta_{m} U_{E}(\textbf{x})
= U_{E}(\textbf{x})-\sum^{\infty}_{n=2}h_{n}u^{tidal}_{nm}(\textbf{x}).
\end{eqnarray}
It represents the variation of Earth gravitational potential due to variation of Earth surface's height in the influence of Moon.

For part (2), we study it by a additional potential $\phi_{m}(\textbf{x})$, which stands for the contribution of redistribution of Earth mass due to Moon's tidal force. Borrowing another Love number $k_{n}$ describing the effect of the Earth mass redistribution in the influence of Moon's tidal force, $\phi_{m}(\textbf{x})$ due to Moon is given by
\begin{eqnarray}\label{12}
  \phi_{m}(\textbf{x})=\sum^{\infty}_{n=2}k_{n}u^{tidal}_{nm}(\textbf{x}).
\end{eqnarray}
According to this equation, the additional potential also has a periodic change in that frequency with the same phase as the periodic variation of the Moon's tidal potential. This contribution has same order of magnitude with contribution of part (1). The Love numbers $h_{n}$ and $k_{n}$ regarded as constant are sufficient for all our calculation on Earth surface. The similar analysis for Moon can be applied to Sun and other celestial bodies.

The total tidal potentials on the Earth surface are the linear superposition of all bodies in the Solar System (excluding the Earth) , which are given by
\begin{eqnarray}\label{4tt}
  u^{tidal}(\textbf{x})=\sum_{b\neq E}u^{tidal}_{b}(\textbf{x}).
\end{eqnarray}
The Earth potential in the influence of the total tidal potentials is rewritten as
\begin{equation}\label{11t}
  U_{E}(\textbf{x}(t))=U_{E}(\textbf{x})+\delta U_{E}(\textbf{x},t)+\phi(\textbf{x},t)
 =U_{E}(\textbf{x})-\!\!\sum^{\infty}_{n=2,b\neq E}(h_{n}-k_{n})u^{tidal}_{nb}(\textbf{x},t),
\end{equation}
which is split into a constant part involving position $\textbf{x}$ and another part involving tidal potentials. It is easy to understand this fluctuating part since it comes from tidal response of the solid Earth.

On the terrestrial surface, the Eq.(\ref{ttransfer1}) is rewritten as
\begin{equation}\label{5t}
\frac{d\tau}{dt}=1-\frac{1}{c^{2}}\Big{[}\frac{v^{2}}{2}+U_{E}(\textbf{x})-\sum^{\infty}_{n=2,b\neq E}(h_{n}-k_{n}-1)u^{tidal}_{nb}(\textbf{x})\Big{]}+O(c^{-4}),
\end{equation}
which demonstrates frequency shift can be split into a constant part and a fluctuating part. For term of $c^{-2}$, the second and third terms compose the constant part, which is determined by the position. The last term is the fluctuating part, which includes tidal response of the solid Earth and astronomical tidal potentials.

The gravitational acceleration is given by $g=\partial_{r}w(\textbf{x}(t))$. Therefore, the change $\delta g$ of gravitational acceleration is given by
\begin{equation}\label{5g}
  \delta g= -\sum^{\infty}_{n=2,b\neq E}\Big{(}1+\frac{2h_{n}}{n}-\frac{(n+1)k_{n}}{n}\Big{)} n\frac{u^{tidal}_{nb}(\textbf{x})}{r}
= -\sum^{\infty}_{n=2,b\neq E}\Big{(}1+\frac{2h_{n}}{n}-\frac{(n+1)k_{n}}{n}\Big{)} \delta g_{n}.
\end{equation}
With this equation, we can easily construct a relation between $\delta g$ and the fluctuating part in frequency shift
\begin{eqnarray}\label{5tg}
  (d\tau/dt)_{f}=-\frac{1}{c^{2}}\sum^{\infty}_{n=2,b\neq E}(1-h_{n}+k_{n})\frac{r}{n}\delta g_{n},
\end{eqnarray}
which demonstrates that the fluctuating part of frequency shift almost is proportional to tidal acceleration. We can apply tides data (for example, gravity tides from superconducting gravimeters) into experiments about the time or frequency. For a reasonable approximation, the Eq.(\ref{5tg}) can be rewritten as a simpler form
\begin{eqnarray}\label{5tgs}
  (d\tau/dt)_{f}=\frac{r(1-h_{2}+k_{2})}{c^{2}(2+2h_{2}-3k_{2})}\delta g.
\end{eqnarray}
The value of Love numbers is given by Earth model. It is sufficient to adopt $h_{2}=0.6$ and $k_{2}=0.3$ for our numerical analysis. In following paragraphs, we will see that the contributions of tidal potentials for clocks in different laboratories are different, which may reach $1\times10^{-17}$ at a distance of $\sim10^{3}$km.

Now, we study the second-order tidal potential. The second-order tidal potential of Moon $u^{tidal}_{2m}(\textbf{x})$ can be written as:
\begin{eqnarray}\label{7}
  u^{tidal}_{2m}(\textbf{x})=\frac{3GM_{m}r^{2}}{4c^{3}_{m}}\Big{[}2\Big{(}\frac{c_{m}}{r_{m}}\Big{)}^{3}\Big{(}\cos^{2}z_{m}-\frac{1}{3}\Big{)}\Big{]},
\end{eqnarray}
where $c_{m}$ is the average distance of the Moon to the center of Earth mass. This formula is the general form for calculating tidal effects. $\cos z_{m}$, $r_{m}$ are the function of time, which we can get by planetary ephemeris.

In order to more conveniently study tidal potential, we rewrite it in the celestial coordinate system (we consider the celestial body is Earth here). We assume that the $\varphi$ is the latitude of observer, $H_{m}$ and $\delta_{m}$ are the local hour angle and declination of the Moon, respectively, which describe Moon's position respect to observer. In the astronomy, the Moon's zenith angle can be expressed as
\begin{eqnarray}\label{8}
  \cos z_{m}=\sin \varphi \sin \delta_{m}+\cos \varphi \cos \delta_{m} \cos H_{m}.
\end{eqnarray}
With the help of this relation, Eq.({\ref{7}}) is rewritten as
\begin{eqnarray}\label{9}
u^{tidal}_{2m}(\textbf{x})=D_{m}\Big{(}\frac{r}{R_{E}}\Big{)}^{2}\Big{(}\frac{c_{m}}{r_{m}}\Big{)}^{3}\Big{[}\cos^{2}\varphi \cos^{2}\delta_{m}\cos2H_{m}\nonumber\\
+\sin 2\varphi \sin2\delta_{m}\cos H_{m}
+3\Big{(}\sin^{2}\varphi-\frac{1}{3}\Big{)}\Big{(}\sin^{2}\delta_{m}-\frac{1}{3}\Big{)}\Big{]},
\end{eqnarray}
where $R_{E}$ is the average radius of Earth, $D_{m}=3GM_{m}R^{2}_{E}/4c_{m}^{3}$ is Doodson constant of Moon \cite{doodson}, whose value is 2.625m$^{2}/$s$^{3}$ from calculation with the Earth parameters of Ref.\cite{12} and astronomy parameters. The first term depends on $\cos 2H_{m}$, which has a period of half a day. The second term depends on $\cos H_{m}$, which has a period of one day. They demonstrate that frequency shift fluctuation has periodic variations of one day and half a day. Then declination $\delta_{m}$ is relevant to the Moon's motion, which modulates frequency shift with a period of half a month. Putting astronomy parameters into this equation, it will give more information. The contribution of Moon's tidal potential for frequency shift is about $10^{-17}$, which is measurable and has a obvious periodicity.
A similar analysis for Sun, it's tidal potential can be expressed by Eq.(\ref{9}) with changing subscript $m$ to $s$.

Then, the third-order tidal potential of Moon $u^{tidal}_{3m}$ may be expressed as
\begin{eqnarray}\label{tid3m1}
   u^{tidal}_{3m}(\textbf{x})=\frac{3GM_{m}r^{2}}{4c^{3}_{m}}\Big{[}\frac{2r}{3c_{m}}\Big{(}\frac{c_{m}}{r_{m}}\Big{)}^{4}(5\cos^{3}z_{m}-3\cos z_{m})\Big{]}.
\end{eqnarray}
This term amounts to several $10^{-19}$ from the fact $r/r_{m}=1/60$. We can express the third-order Sun tidal potential as the same form of Eq.(\ref{tid3m1}). Since $r/r_{s}=1/234000$, $u^{tidal}_{3s}$ is neglected. Inserting Eq.(\ref{8}) into Eq.(\ref{tid3m1}), we obtain
\begin{equation}\label{tid3m2}
  u^{tidal}_{3m}(\textbf{x})= D_{m}\Big{(}\frac{r}{R_{E}}\Big{)}^{3}\Big{(}\frac{c_{m}}{r_{m}}\Big{)}^{4}\Big{(}\frac{R_{E}}{c_{m}}\Big{)}(A_{0}+A_{1}\cos H_{m}
  +A_{2}\cos 2H_{m}+A_{3}\cos 3H_{m}),
\end{equation}
where the previous three terms have same periods with the second-order tidal potential and the fourth term is related to $\cos 3H_{m}$, which has a period of 1/3 day. The other parameters $A_{i}(i=0...3)$ are position-related functions.

\section{Perturbed Kepler orbit}\label{appa}
In this appendix, we introduce perturbed Kepler problem. The motion of a satellite under the gravitational attraction of a spherical body can be solved by the Keplerian orbit. However, the actual orbit of satellite deviates from Kepler's. The motion of satellite is perturbed by the oblateness of the Earth. The gravitational attractions of Sun, Moon and other planets also can change the motion of satellite in the vicinity of the Earth. Under the influences of these factors, the orbital equation of satellite cannot be solved exactly and completely. While these systems cannot be given an exact description, we can make progress by approximate method. The additional forces have only a small effect on the motion of satellite around the Earth. The gravitational effects of Sun and Moon are main external influences on the Earth-satellite system. The approximate method of external gravitational influences on the satellite's motion is the realm of perturbation theory. The well-known method is $\emph{method of osculating orbital elements}$ \cite{bookg1}. The method is to calculate variation of orbital elements by integrating differential equations. So, the perturbed Kepler problem can be split into two parts: orbital elements of Keplerian orbit and the change of orbital elements due to perturbing forces.

The Keplerian orbit may be given by six orbital elements: semi-major axis $a$, the eccentricity $e$, the true anomaly $f$, the inclination of the orbital plane to the reference plane $i$, the longitude of pericenter $\omega$ and the longitude of the ascending node $\Omega$. In the orbital plane, the Keplerian orbit is given by
$r=p(1+e\cos{f})^{-1}$,
where $p=a(1-e^{2})$ is the orbit's semi-latus rectum and can be used as an element instead of semi-major axis $a$. We assume that orbital frame is given by coordinate system $(x,y,z)$, further orbital plane coincides with $x-y$ plane, and perturbing force frame is given by coordinate system $(X,Y,Z)$, further perturbing force is in the $X-Y$ plane.

We consider that the small perturbing force $\textbf{f}$ on the satellite is decomposed as
\begin{eqnarray}\label{perf}
  \textbf{f}=\mathcal{A}\textbf{n}+\mathcal{B}\textbf{n}_{\bot}+\mathcal{C}\textbf{e}_{z}.
\end{eqnarray}
By the method of osculating orbital elements, the secular variation of the Keplerian orbital elements are given by following equations\cite{bookg1}:
\begin{eqnarray}\label{deldt}
  \frac{da}{dt}=2\sqrt{\frac{a^{3}}{Gm}}(1-e^{2})^{-1/2}[e\sin{f}\mathcal{A}+(1+e\cos{f})\mathcal{B}],\\
  \frac{de}{dt}=\sqrt{\frac{p}{Gm}}\Big{[}\sin{f}\mathcal{A}+\frac{2\cos{f}+e(1+\cos^{2}f)}{1+e\cos{f}}\mathcal{B} \Big{]},\\
  \frac{df}{dt}=\sqrt{\frac{Gm}{p^{3}}}(1+e\cos{f})^{2}+\frac{1}{e}\sqrt{\frac{p}{Gm}}\Big{[}\cos{f}\mathcal{A}-\frac{2+e\cos{f}}{1+e\cos{f}}\sin{f}\mathcal{B} \Big{]},
\end{eqnarray}
\begin{eqnarray}\label{deldtit}
   \frac{dp}{dt}=2\sqrt{\frac{p^{3}}{Gm}}\frac{1}{1+e\cos{f}}\mathcal{B}.
\end{eqnarray}
Considering more detailed description of orbit, it is indispensable to take other orbital elements into account.  With this formalism, we can calculate the variation of orbital elements with the iterative integral of these equations.

For calculating the tidal effects of clock, we consider an example where the perturbing force is tidal force from Moon. The orbital plane has an inclination $i$ relative to the orbital plane of Moon that coincides with $X-Y$  plane. The tidal force due to Moon may be expressed by expanding in powers of $r/r_{m}$, and the leading term is
\begin{eqnarray}\label{tidalfm}
  \textbf{f}=-\frac{GM_{m}r}{r_{m}^{3}}[\textbf{n}-3(\textbf{n}\cdot\textbf{n}_{m})\textbf{n}_{m}].
\end{eqnarray}
In the case of $r\ll r_{m}$, the tidal force is a small perturbing force comparing Earth's gravity. For low-orbit satellites around Earth, it is sufficient to use the $\emph{method of osculating orbital elements}$.
In Eq.(\ref{tidalfm}), we assume that the direction from Earth to Moon is
\begin{eqnarray}\label{emvec}
  \textbf{n}_{m}=\cos{F}\textbf{e}_{X}+\sin{F}\textbf{e}_{Y},
\end{eqnarray}
where $F=F_{0}+\Omega_{m}t$ is the true anomaly of Moon, $\Omega_{m}$ is the angular frequency of Moon orbit. Furthermore, from the coordinate transformation between coordinate system $(x,y,z)$ and coordinate system $(X,Y,Z)$, the orbital basis vectors $(\textbf{n},\textbf{n}_{\bot},\textbf{e}_{z})$ are expressed as
\begin{eqnarray}\label{orbvec}
\textbf{n}=&&[\cos\Omega\cos{(f+\omega)}-\cos{i}\sin\Omega\sin{(f+\omega)}]\textbf{e}_{X}+\nonumber\\
           &&[\sin\Omega\cos{(f+\omega)}+\cos{i}\cos\Omega\sin{(f+\omega)}]\textbf{e}_{Y}
           +\sin{i}\sin{(f+\omega)}\textbf{e}_{Z},\nonumber\\
\textbf{n}_{\bot}=&&[-\cos\Omega\sin{(f+\omega)}-\cos{i}\sin\Omega\cos{(f+\omega)}]\textbf{e}_{X}+\nonumber\\
                  &&[-\sin\Omega\sin{(f+\omega)}+\cos{i}\cos\Omega\cos{(f+\omega)}]\textbf{e}_{Y}
                   +\sin{i}\cos{(f+\omega)}\textbf{e}_{Z},\nonumber\\
\textbf{e}_{z}=&&\sin{i}\sin\Omega\textbf{e}_{X}-\sin{i}\cos\Omega\textbf{e}_{Y} +\cos{i}\textbf{e}_{Z}.
\end{eqnarray}
Inserting Eqs. (\ref{emvec})(\ref{orbvec}) into (\ref{tidalfm}), the components of tidal force are
\begin{equation}\label{tkpa}
  \mathcal{A}=-\frac{GM_{m}r}{r^{3}_{m}}[1-3(\cos{(f+\omega)}\cos{(F-\Omega)}+\sin{(f+\omega)}\sin{(F-\Omega)}\cos{i})^{2}],
\end{equation}

\begin{eqnarray}\label{tkpb}
 \!\!\!\!\!\!\!\!\!\!\!\!\!\!\!\!\!\!\!\!\!\!\!\!\!\!\!\!\!\!\! \mathcal{B}=-\frac{3GM_{m}r}{r^{3}_{m}}[\cos{(f+\omega)}\cos{(F-\Omega)}+\sin{(f+\omega)}\sin{(F-\Omega)}\cos{i}]\nonumber\\
\times[\sin{(f+\omega)\cos{(F-\Omega)}}-\cos{(f+\omega)}\sin{(F-\Omega)}\cos{i}],
\end{eqnarray}

\begin{equation}\label{tkpc}
 \mathcal{C}=-\frac{3GM_{m}r}{r^{3}_{m}}[\cos{(f+\omega)}\cos{(F-\Omega)}
  +\sin{(f+\omega)}\sin{(F-\Omega)}\cos{i}]\sin{(F-\Omega)}\sin{i},
\end{equation}
Inserting these components into Eqs.(\ref{deldt})-(\ref{deldtit}), we can calculate the variation of orbital elements due to Moon's tidal force.

As an example, we assume that satellite and Moon move in the same orbital plane(it implies inclination $i=0$). 
For the low-orbit satellite around Earth, it's orbital period is far smaller than Moon's. Then, the true anomaly of Moon $F$ change very little for a complete satellite orbit and  $F$ can be regarded as a constant over a complete orbit.
Satellite's orbital evolution in response to Moon is governed by Eqs. (\ref{deldt})-(\ref{deldtit}).  After a complete satellite orbit, the net change of orbital elements is
\begin{eqnarray}\label{elementd1}
  \Delta{p}=-15\pi\frac{M_{m}p^{4}}{M_{E}r^{3}_{m}}\frac{e^{2}}{\sqrt[7]{1-e^{2}}}\sin{2(\omega-F+\Omega)},
\end{eqnarray}
\begin{eqnarray}\label{elementd2}
  \Delta{e}=\frac{15\pi M_{m}p^{3}}{2M_{E}r^{3}_{m}}\frac{e}{\sqrt[5]{1-e^{2}}}\sin{2(\omega-F+\Omega)},
\end{eqnarray}
and the change of other elements is zero ($\Delta\omega$ is not zero in this case, however its variation doesn't affect our calculations, we ignore it here). The similar procedure may be used to calculate the variation of orbital elements in the influence of Sun. In addition, we should add high-order terms in the expansion of tidal force in powers of $r/r_{m}$ for high-orbit satellites.

\end{document}